\documentclass[a4paper,10pt]{article}

\usepackage{amssymb,amsfonts,amsmath,epsfig,float,graphicx}
\usepackage{xcolor}

\usepackage[dvips]{hyperref}

\title{Emergent Coulomb forces in reducible Quantum Electrodynamics}
\author{Jan Naudts\\
Universiteit Antwerpen\\
\url{Jan.Naudts@uantwerpen.be}\\
\url{https://orcid.org/0000-0002-4646-1190}
}
\date{}

\def\kk{{\bf k}}
\def\Ro{{\mathbb R}}
\newcommand{\be}{\begin{eqnarray}}
\newcommand{\ee}{\end{eqnarray}}
\newcommand{\nnee}{\nonumber\ee}

\newcommand{\upd}{{\rm d}}
\def\kkph{{\kk^{\mbox{\tiny ph}}}}
\def\kk{{\bf k}}

\newtheorem{theorem}{Theorem}[section]

\newtheorem{proposition}[theorem]{Proposition}

\newtheorem{definition}[theorem]{Definition}
\def\beginproof{\par\strut\vskip 0.1cm\noindent{\bf Proof}\par}
\def\endproof{\par\strut\hfill$\square$\par\vskip 0.5cm}

\newcommand{\Gamman}{\Gamma_{\mbox{\tiny \strut\hskip -2pt norm}}}

\newcommand{\Gammadual}{\Gamma^\dagger}
\newcommand{\Gammaphel}{\Gamma^{\mbox{\tiny ph,el}}}

\newcommand{\Rotn}{\Ro^3_{\mbox{\tiny o}}}

\def\hc{^\dagger}
\def\trans{^{\mbox{\tiny T}}}
\newcommand{\Tr}{\,{\rm Tr}\,}
\def\Cc{C_{\mbox{\tiny c}}}

\def\Co{{\mathbb C}}

\def\Io{{\mathbb I}}

\def\Ro{{\mathbb R}}

\def\eev{{\bf e}}

\def\kk{{\bf k}}

\def\xx{{\bf x}}
\def\yy{{\bf y}}

\def\kkp{{\bf k'}}
\def\kkpp{{\bf k''}}
\def\kkph{{\kk^{\mbox{\tiny ph}}}}
\def\kph{k^{\mbox{\tiny ph}}}

\def\uc{q_{\mbox{\tiny el}}}

\def\phicl{\phi^{\mbox{\tiny cl}}}
\def\lu{{\ell}}
\def\mstar{\kappa}

\def\ranglec{\rangle^{\mbox{\tiny c}}}
\def\langlec{\strut^{\mbox{\tiny c}}\hskip -1pt\langle}

\def\Acl{A^{\mbox{\tiny cl}}}
\def\Ecl{E^{\mbox{\tiny cl}}}
\def\Bcl{B^{\mbox{\tiny cl}}}
\def\ah{a_{\mbox{\tiny H}}}
\def\av{a_{\mbox{\tiny V}}}
\def\ahdagger{a\hc_{\mbox{\tiny H}}}
\def\avdagger{a\hc_{\mbox{\tiny V}}}

\def\Hph{H^{{\mbox{\tiny ph}}}}
\def\HI{H^{{\mbox{\tiny I}}}}
\def\hatHI{\hat H^{{\mbox{\tiny I}}}}

\def\Hel{H^{{\mbox{\tiny el}}}}
\def\Hcalem{{\cal H}_{\mbox{\tiny em}}}
\def\Hqu{{\cal E}^{\mbox{\tiny qu}}}
\def\Hcl{{\cal E}^{\mbox{\tiny cl}}}
\def\Nph{N_0}
\def\Nel{N_\mstar}

\def\psicha{\psi^{\mbox{\tiny a}}}
\def\pc{r}
\def\Pc{R}
\def\Ga{G}

\def\calEph{{\cal E}^{\mbox{\tiny ph}}}
\def\calEel{{\cal E}^{\mbox{\tiny el}}}
\def\calEint{{\cal E}^{\mbox{\tiny int}}}
\def\rhoel{\rho^{\mbox{\tiny el}}}
\def\rhoph{\rho^{\mbox{\tiny ph}}}
\def\rhovac{\rho^{\mbox{\tiny vac}}}

\begin{document}
\maketitle

\begin{abstract}
This paper discusses an attempt to develop a mathematically rigorous theory
of Quantum Electrodynamics (QED). It deviates from the standard version of QED
mainly in two aspects: it is assumed that the Coulomb forces are carried by
transversely polarized photons, and a reducible representation of the
canonical commutation and anti-commutation relations is used. 
Both interventions together should suffice to eliminate the mathematical
inconsistencies of standard QED. 

\end{abstract}

\section{Introduction}

In recent work \cite{VEP11,VEP16} Erik Verlinde formulated the claim that gravity is an emergent force.
By this is meant that gravitational forces can be derived from other, more fundamental forces.
Coulomb forces and gravitational forces have much in common. Both are inversely proportional to the
square of the distance. And it is difficult to reconcile them with the aversion of physicists to action at a distance.
It is therefore obvious to claim that also the Coulomb forces are emergent forces \cite{NJ17}.
While the claim of Verlinde is made in the context of cosmology the emergence of the Coulomb forces
is formulated here in the context of quantum field theory.

Two arguments are brought forward. In the common theory of Quantum Electrodynamics (QED), which includes
Coulomb forces, it is possible to remove them by a simple transformation of observables \cite{CM79}.
The inverse transformation can be used to reintroduce Coulomb forces given a theory in which they are
absent (See Appendix \ref {app:emerg}).
The other argument is based on a mathematical proof that electrons bind with long-wavelength photons. This mechanism is discussed further on in the present paper.
Evidence for the feasibility of this kind of binding has been reported recently \cite{SZKBSMBSV18}
in experiments involving single spins in silicon quantum dots
binding with long-wavelength microcavity photons.

A theory of Quantum Electrodynamics which does not include Coulomb forces is appealing
because it can avoid some of the technical problems which plague the standard
version of the theory. The claim made in  \cite{NJ17} is that Coulomb attraction and repulsion
are carried by transversely polarized photons. This eliminates the need for longitudinal
and scalar photons and allows for building a rigorous theory of QED.
In the present paper the mathematical arguments supporting this claim are presented in appendix.

\section{The Standard Model}

The Standard Model of elementary particles summarizes much of our present day understanding
of the fundamental laws of physics. It is a highly effective theory. It explains almost all
phenomena with an amazing degree of precision. Never the less, it cannot be the final theory of physics,
in the first place because the integration with Einstein's general relativity theory is missing.

The Standard Model is a quantum field theory, which implies that particles are
described by fields, for instance a photon field or an electron field.
Creation and annihilation operators add or remove one quantum of the corresponding field.
This wording suggests that particles are created or annihilated.
However, it is more careful to say that creation and annihilation operators
are a tool to construct quantum fields and to describe interaction processes.

Two different fields can interact with each other by the exchange of one or more quanta.
Einstein proposed this mechanism to explain the photo-elec\-tric effect.
This assumption is a corner stone of all quantum theories. The interesting question
why the exchange of quanta occurs at seemingly random discrete moments of time
is not discussed here.

The prototype of a quantum field theory is Quantum Electrodynamics (QED),
the relativistic quantum theory of electromagnetism.
QED was extended to include all electro-weak interactions, and later on
also strong interactions. Much effort goes into the exploration of further extensions. 
The whole construction has become an inverted pyramid resting on a few basic
principles most of which were decided on in the early days of quantum mechanics.
It is therefore of utmost importance that existing holes in these foundations
are eliminated. 

It is indeed worrying that the common formulation
of QED is mathematically inconsistent. A first reason for that is the perturbative approach,
which involves a non-converging series expansion. It is even worse: many of the individual
terms of the expansion are ill-defined. As a consequence, repair techniques are needed. 
An alternative is offered by non-perturbative QED. But even then not all problems disappear.

\section{Mathematical problems}

The technical difficulties of QED have their consequences. Certain questions
cannot be treated in a reliable manner. Let me mention one.

The physical vacuum differs from the mathematical vacuum. The latter is defined
by a vanishing of all fields. However it is not the state of minimal energy.
The interaction between different fields creates states of negative energy.
A well-known example in the quantum mechanics of particles is the hydrogen atom. Its eigenstates
all have negative energy while the scattering states have positive energy.
By analogy one can expect that in QED there exist states of negative energy in which
the electromagnetic field and the electron/positron field are bound in some way.
If this expectation is correct then the state of minimal energy is the physical vacuum.
No mathematical proof of its existence exists. In a non-relativistic context Lieb and
coworkers (See for instance \cite {LS09}) prove the existence of a ground state for models
consisting of particles which do not interact among themselves but do interact with the radiation field.
An ultra-violet cutoff is applied to the latter.

In Quantum Chromodynamics (QCD) one accepts that
a symmetry-breaking phase transition occurs at low energies.
Such a phase transition gives a satisfactory explanation for the confinement phenomenon.
Also the Englert-Brout-Higgs mechanism, which in the Standard Model is responsible
for giving mass to particles, invokes a phase transition. Because these phase transitions
are continuous and symmetry-breaking the physical ground state is non-unique and is
accompanied by Goldstone bosons. Mathematical proofs for these statements are missing.

\section{Reducible QED}

One way out to aim at a rigorous theory of Quantum Electrodynamics
is to give up one of the axioms of quantum field theory.
In the context of constructive quantum field theory \cite{SW64}
evidence exists that in Minkowski space the only models
satisfying the Wightman axioms are free-field models.
Hence some modification of the underlying assumptions is indicated.

A mild modification of the accepted body of axioms is to
allow for reducible representations of the Lie algebra
of canonical commutation and anti-commutation relations.
Non-commuting observables are at the heart of quantum
mechanics, while in classical mechanics all observables
mutually commute. The two theories, classical and quantum,
can be made as far apart as possible by requiring that
in quantum mechanics the only observables which
commute with all others are the multiplications with a scalar number.
If this is the case then the representation of the algebra of observables is
said to be irreducible. In addition, any representation can be decomposed
into irreducible representations \cite{DJ69}. However, the argument that
it therefore suffices to study the irreducible representations
is misleading. This may be so when the decomposition is
discrete and involves a small number of irreducible
components. The continuous decompositions considered in what follows
add a degree of complexity rather than simplify the theory.

Reducible QED is studied in the work of Czachor and collaborators
(See \cite{CM00,CM04,CN06,CW09} and references given in these papers).
The main assumption is that the irreducible representations
are labeled by a three-dimensional wave vector $\kk$
and that the decomposition of the reducible representation
is then an integral over the wave vector $\kk$.
The version presented here differs from the original
version of Czachor by making explicit that for every wave vector
$\kk$ there is a properly normalized wave function $\zeta_\kk$
and for every observable $\hat A$ there is a local copy $A_\kk$
such that the quantum expectation of $A$ is given by
\be
\langle \hat A\rangle&=&\int\upd\kk\,\langle\zeta_\kk|A_\kk\zeta_\kk\rangle.
\nonumber\ee
For additive quantities such as the total energy this expression
is quite obvious: the total energy of the field is obtained by
integrating the wave vector-dependent energy density.
Details of this formalism are found in Appendix \ref {app:RQF}.

Reducible representations are known in quantum field theory in
the context of superselection rules. In the latter case one selects
a single component $\zeta_\kk$ which represents the state of the system.
Here the quantum field is represented by all $\zeta_\kk$ simultaneously.
The selected wave vector $\kk$ is just a filter through which
the quantum field is looked at.

\section{World view}

Historically, the first picture one had of Quantum Electrodynamics
was that of ordinary space filled with quantum harmonic oscillators,
one pair at each point to cope with the two polarizations of the free
electromagnetic field. This picture survives here with the modification
that there is one pair of quantum oscillators for each wave vector $\kk$.
Hence, the Euclidean space is replaced by its Fourier space.
The 3 classical dimensions appear because the representation of
the two-dimensional quantum harmonic oscillator is not the
irreducible one but is reducible.

The representation of fermionic fields such as that of electrons and positrons
is also reducible. The reducing wave vector is independent of the photonic wave vector.
The Hilbert space of the irreducible fermionic representation is finite-dimensional.
The electron/positron field has 16 independent states for each wave vector $\kk$.
Its Fock space is obtained from the vacuum state by the action of 4 fermionic
creation operators adding an electron or a positron, each with either spin up or spin down.

An important difference between reducible QED and QED of the Standard Model
concerns the superposition of photons with unequal wave vector.
In the present version of reducible QED this superposition
requires another field as intermediary. For instance, expression
(13) of \cite{HZ86}
\be
\frac 1{\sqrt 2}\left[|-K\rangle_1\,|K\rangle_2
+|-K'\rangle_1\,|K'\rangle_2\right]
\nonumber\ee
becomes here
\be
\zeta_{-\kk}=|-\kk,\uparrow\rangle_1,
&\quad&
\zeta_{\kk}=|\kk,\uparrow\rangle_2,\cr
\zeta_{-\kk'}=|-\kk',\downarrow\rangle_1,
&\quad&
\zeta_{\kk'}=|\kk',\downarrow\rangle_2.
\nonumber\ee
Here, $|\uparrow\rangle$ and $|\downarrow\rangle$ are for instance the spin up respectively
spin down states of an auxiliary electron field.
The wave function $\zeta$ describes the simultaneous presence of two combinations.
The former describes two photons with wave vector $\pm \kk$
combined with an electron with spin up, the latter describes two photons with wave vector $\pm \kk'$
combined with an electron with spin down.
In conclusion: with a single field superpositions can be made at a given wave vector.
Wave functions at different values of the wave vector are always present
and do no require further superposition. Entangled states require two or more distinct
fields.

Many expressions in reducible QED look similar to those of common QED 
except that the integration over wave vectors is missing. It is shifted
towards the evaluation of expectation values. Commutation and anti-commutation relations
look different because Dirac delta functions are replaced by cosine and sine functions.
An immediate advantage is that operator-valued distribution functions are avoided
and that problems caused by ultra-violet divergences are postponed.

In fact, it is not clear what happens with the electromagnetic field at large energy density. 
High energy fields can be
obtained in two ways, either by increasing the energy of individual photons or
by increasing the number of photons. Recent experiments are exploring the
situation (See for instance \cite {PMHK12}).

\section{The radiation gauge}

In the present version of reducible QED only transversely polarized photons occur
(See Appendix \ref{sect:scalbos}, \ref {section:emf}).
The longitudinal and scalar photons of the traditional theory are absent.
This implies that the number of degrees of freedom of the electromagnetic field is 2
rather than 3 or 4. There is no need for the construction of Gupta \cite {GS50} and Bleuler \cite{BK50},
which intends to remove the nonphysical degrees of freedom. This is an important simplification,
which however raises a number of questions.

In a theory containing only transverse photons it is obvious to use the radiation gauge.
This is the Coulomb gauge \cite {NoteCoulombGauge},
which is often used in Solid State Physics, in absence of
Coulomb forces. A drawback of using this gauge is that it is not manifestly Lorentz covariant.
What one wins by using this simplifying gauge is lost at the moment one considers
a Lorentz boost. Then calculations, needed to restore the radiation gauge, are rather painful.
However, this is not a fundamental problem.

If no gauge freedom is left, what is then the role of gauge theories? They are 
the unifying concept behind the different boson fields appearing in the Standard Model. 
The reasoning goes that a global
gauge symmetry of the free fields becomes a local symmetry of the interacting fields.
Total charge $Q$ is a conserved quantity also in reducible QED.
The corresponding symmetry group of unitary operators $e^{i\Lambda Q}$
corresponds with the U(1) gauge group of the Standard
Model. It multiplies the wave function of an electron with a phase factor $e^{i\Lambda q}$.
It is shown in Appendix  \ref {app:gauge}
that also in reducible QED this global symmetry group can be extended with
local symmetries, where 'local' now means local in the space of wave vectors.
The constant $\Lambda$ then becomes a function of the wave vectors $\kkph$ of the photon and $\kk$
of the electron/positron field. It suffices that $\Lambda$ remains constant when
$\kk$ is replace by $\pm\kk\pm\kkph$. Then $e^{i\Lambda Q}$ is still a symmetry of the Hamiltonian.
It reflects that the wave vector of the electron/positron field can only change by emission or
absorption of a photon.

\section{Emergence of Coulomb forces}

The main concern for a theory involving only transverse photons is how to explain
the Coulomb forces observed in nature. The explanation given in the present theory
is based on an analogy with long range forces which act between polarons.
The polaron \cite{DA09} is a concept of Solid State Physics.
A free electron in a dielectric crystal interacts with lattice vibrations,
called phonons, and can form a bound state with them. This bound state is the polaron.
Two polarons interact with each other because they share the same lattice vibrations.
This interaction is long-ranged.

Similarly, an electron of QED interacts with 
the electromagnetic field.
The distinction between a hypothetical {\em bare} electron
and a {\em dressed} electron, surrounded by a cloud of photons, is as old as
QED itself. The effect of the electron on itself via its interaction with the electromagnetic
field yields a contribution to its energy. This is called the {\em self-energy} of the electron.
In an electrostatic context the self-energy is the potential energy of the
charge of the electron in its own Coulomb field. For a point particle this
contribution is infinite. However, in the present paper the Coulomb field is absent by
assumption. Hence, the problem of this divergent energy, in its original form, disappears.
The interaction between the electron/positron field and the electromagnetic field
still yields a static contribution to the energy. As discussed below it decreases
the total energy instead of blowing it up to +infinity. In addition, the 
effects of the dynamic interaction between the two fields are very complicated.
For a discussion of the latter see for instance \cite{PMHK12}.

In reducible QED one can prove
that an electron can form a bound state with a
transversely polarized photon field (See Appendix \ref {app:boundstate}).
The binding energy is minus twice the kinetic energy
of the photon field. This result is typical for a linear interaction of the electron field with
a photon field described by a quadratic Hamiltonian. In some cases the self-energy
is negative (See Appendix \ref{app:var}).
The bound state formed in this way is similar to the polaron.
In addition, Appendix \ref {app:longwave}
shows that the binding also exists for photons with long wavelength
and with a wave vector almost parallel to that of the electron field.
This is important because it makes it plausible that different regions of the electron field
develop a long-range interaction, which is then the Coulomb interaction.
However, a mathematical proof that the Coulomb attraction and repulsion are reproduced by this mechanism
is still missing.

\section{Discussion}

The assumption that the Coulomb forces are emergent instead
of being fundamental forces has far reaching consequences,
some of which have been discussed above.
A number of problems of the standard theory are eliminated or can at least be avoided.
No superfluous degrees of freedom appear
because only transversely polarized photons are taken into account.
The problem of the divergence of the self-energy of the electron
is absent. 

The combination of reducible QED with the assumption of emergent Coulomb forces
allows for the development of a mathematically consistent theory of QED.
In particular, ultraviolet divergences are not hindering because the integration
over wave vectors is postponed to the moment of evaluation of quantum expectations.
In this context rigorous proofs can be given of the existence of bound states
due to the interaction of the electron field with long wave length photons.
An analogy with the polarons of Solid State Physics makes it then plausible that
Coulomb forces are carried by these transversely polarized long wavelength photons.

Some experimental evidence for the present version of reducible QED
is found in Solid State Physics.
It is generally accepted that free electrons in metals do not experience
any long range Coulomb repulsion. In the present context
this is an immediate consequence of the characteristic
property of metals that long wavelength photons cannot propagate inside the material.
Evidence for the binding of the electron spin with long-range photons is given in \cite{SZKBSMBSV18}.
A prediction of the present version of reducible QED is that 
entanglement of photons with distinct wave vectors requires an ancillary field.
It is not easy to test this property because in any experiment entanglement with
the environment is hard to avoid.

The claim that the Coulomb forces are emergent, if correct, requires significant
modifications to the Standard Model. The concept of gauge theories
survives in a modified form, as indicated. The gauge freedom is not any longer
due to the presence of superfluous degrees of freedom but is
the consequence of working in a reducible representation.
What this means for the weak and strong interactions has still to be investigated.

It is tempting to extrapolate the present work in the direction of quantum gravity.
Technical difficulties seem treatable. The existence of long wavelength gravitational
waves has been established recently \cite{ABP}. However, the existence of the graviton as the quantum
of the gravitational field is still an open question. It is also hard to believe
that the gravitational forces, which we experience all the time, would be carried by
low energy quantum particles about which we do not know anything.

\appendix
\section*{Appendices}
\section{Reducible quantum fields}
\label{app:RQF}
\subsection{Definition}

Let $\cal H$ is a given Hilbert space,
either finite dimensional or separable, and $K$ an open subset of $\Ro^n$.
In the sequel $K$ will be either $\Ro^n$ or $\Ro^n\setminus \{0\}$.
Normalized elements of $\cal H$ are called {\em wave functions}, elements of $K$
are called {\em wave vectors}.
Maps of $K$ into $\cal H$ are called {\em quantum fields}.

Let $\Gamma$ denote the linear space of continuous fields
\be
\zeta:\,\kk\in K\mapsto \zeta_\kk\in {\cal H}.
\nnee
In the terminology of \cite{DJ69} $\Gamma$ is a {\em continuous field of Hilbert spaces}.
A family of sesquilinear forms $(\cdot,\cdot)_\kk$, $\kk\in K$, is defined  on $\Gamma$ by
\be
(\phi,\zeta)_\kk=(\phi_\kk,\zeta_\kk).
\nnee
The corresponding semi-norms $||\zeta||_\kk\equiv||\zeta_\kk||$
turn $\Gamma$ into a locally convex Hausdorff space.

A subspace $\Gamman$ of $\Gamma$ is formed by the $\zeta\in\Gamma$
for which the map $\kk\mapsto ||\zeta_\kk||$ is bounded continuous.
A norm is defined on this subspace by
\be
||\zeta||&=&\sup_{\kk\in K}||\zeta_\kk||.
\nnee
It turns $\Gamman$ into a Banach space.
Fields belonging to this subspace are said to be bounded in norm.

In standard quantum mechanics the normalization of wave functions is important.
In the present context this leads to the axiom that
{\em states} of the quantum field theory are represented
by elements $\zeta$ of $\Gamma$ which satisfy the normalization condition
\be
||\zeta_\kk||=1\quad\mbox{ for all }\kk\in K.
\nnee
If this condition is satisfied then $\zeta\in\Gamma$ is said to be {\em properly normalized.}

\subsection{Transposed fields}

The dual $\Gamma^*$ of $\Gamma$ consists of all continuous conjugate-linear functions of
$\Gamma$. Introduce

\begin{definition}
 A dual field $\theta$ is a map $\kk\in K\mapsto \theta_\kk\in\Gamma^*$ which is point-wise continuous.
\end{definition}

\noindent
The space of dual fields is denoted $\Gammadual$.
Introduce also the notation
\be
(\theta,\zeta)_\kk&=&\overline{\theta_\kk(\zeta)}, \quad\zeta\in\Gamma, \theta\in\Gammadual.
\nnee
Because $\zeta\mapsto \theta_\kk(\zeta)$ is conjugate-linear the form $(\cdot,\cdot)$ is sesquilinear.
Following the Physics convention it is linear in the second argument.
The requirement of point-wise continuity in the definition means that the map
$\kk\mapsto (\theta,\zeta)_\kk$ is continuous for any field $\zeta$ in $\Gamma$.

Given $\theta\in\Gamma$ let $\theta\trans_\kk$ be defined by
\be
\theta\trans_\kk:\zeta\in\Gamma\mapsto \langle\zeta_\kk|\theta_\kk\rangle.
\nnee
It belongs to $\Gamma^*$ and the map
$\theta\trans:\,\kk\mapsto \theta\trans_\kk$
is a dual field. This shows that $\Gamma$ is embedded in the set of
dual fields by the injection $\theta\mapsto \theta\trans$.
One has
\be
(\theta\trans,\zeta)_\kk
&=&\overline{\theta\trans_\kk(\zeta)}\cr
&=&\overline{\langle\zeta_\kk|\theta_\kk\rangle}\cr
&=&\langle\theta_\kk|\zeta_\kk\rangle
\nnee
and
\be
(\theta\trans,\zeta)_\kk=\overline{(\zeta\trans,\theta)_\kk}.
\nnee
The space of transposed fields $\theta\trans$, $\theta\in\Gamma$
is denoted $\Gamma\trans$ and is a subspace of $\Gammadual$.
The inverse transposition is the map $\theta\trans\mapsto\theta$.
It is tradition to call this inverse map also a transposition
and to convene that $(\theta\trans)\trans=\theta$.

\subsection{Diagonal operators}

A linear operator $\hat A$ in $\Gamma$ is a {\em diagonal operator} if there exists
a map $\kk\in K\mapsto A_\kk$, where $A_\kk$ is a linear operator on $\cal H$,
and a subspace $\cal D$ of $\Gamma$, called the domain of $\hat A$,
such that for all $\zeta$ in $\cal D$
\begin{itemize}
 \item $\zeta_\kk$ is in the domain of $A_\kk$ for all $\kk$;
 \item $\kk\mapsto A_\kk\zeta_\kk$ is continuous;
 \item $\hat A\zeta$ equals the map $\kk\mapsto A_\kk\zeta_\kk$.
\end{itemize}

The diagonal operators generalize the concept of block-diagonal matrices
for which all blocks have the same size. In fact, if $K$ is a finite set
and $\cal H$ is finite-dimensional then any diagonal operator is represented
by a block-diagonal matrix.

Any operator $A$ on $\cal H$ defines a diagonal operator $\hat A$ on $\Gamma$ by
\be
[\hat A\zeta]_\kk&=&A\zeta_\kk\quad\mbox{ for all }\kk\in K.
\nnee
The domain of this operator is the set
\be
{\cal D}&=&\{\zeta\in\Gamma:\, \zeta_\kk\mbox{ is in the domain of }A\mbox{ for all }\kk\in K\}.
\nnee
In particular, the identity operator $\Io$ is a diagonal operator
which satisfies $\hat\Io\zeta=\zeta$ for all $\zeta\in\Gamma$.

\begin{proposition}
If $A$ is a bounded operator on $\cal H$ then
\begin{description}
 \item 1) $\hat A$ is a continuous operator
defined on all of $\Gamma$;
\item 2) If $\zeta\in\Gamma$ is bounded in norm then also $\hat A\zeta$ is bounded in norm
and $||\hat A||=||A||$.
\end{description}
\end{proposition}

\beginproof
\paragraph{1)}
For any $\zeta\in\Gamma$ is
\be
||A\zeta_\kk-A\zeta_\kkp||
&\le&
||A||\,||\zeta_\kk-\zeta_\kkp||.
\nnee
Hence, continuity of $\kk\mapsto A\zeta_\kk$ follows from the continuity of
$\kk\mapsto \zeta_\kk$.
This shows that any $\zeta$ in $\Gamma$ belongs to the domain of $\hat A$.
Finally, continuity of $\hat A$ follows because it suffices that for each $\kk$
the seminorm $||\hat A\zeta||_\kk$ is bounded above by $||A||\,||\zeta||_\kk$.

\paragraph{2)}
If $\zeta\in\Gamma$ is bounded in norm then
\be
||\hat A\zeta||
&=&\sup_\kk ||[\hat A\zeta]_\kk||\cr
&=&\sup_\kk ||A\zeta_\kk||\cr
&\le&||A||\,\sup_\kk||\zeta_\kk||\cr
&=&
||A||\,||\zeta||.
\nnee
Hence $\kk\mapsto A\zeta_\kk$ is bounded in norm and $||\hat A||\le||A||$.
Equality $||\hat A||=||A||$ follows from the action of $\hat A$ on constant fields.

\endproof

\begin{proposition}
If $A$ is a closed operator on $\cal H$ then $\hat A$ is a closed operator on
the Banach space $\Gamman$.
\end{proposition}

\beginproof

Assume $\zeta^{(n)}$ converge in norm to $\zeta$ and $\eta^{(n)}\equiv \hat A\zeta^{(n)}$
converge to $\eta$.
Then for each $\kk\in K$ converges $\zeta^{(n)}_\kk$ to $\zeta_\kk$ and
$\eta^{(n)}_\kk=A\zeta^{(n)}$ converge to $\eta_\kk$.
Because $A$ is closed with given domain ${\cal D}_A\subset{\cal H}$ the vector $\zeta_\kk$
belongs to ${\cal D}_A$ and $A\zeta_\kk=\eta_\kk$. 
Because $\eta\in \Gamman$ the map $\kk\mapsto\eta_k$ is continuous. Hence,
$\zeta$ belongs to $\cal D$ and $\hat A\zeta=\eta$.
\endproof

An example of diagonal operators is found in the book of Dixmier \cite{DJ69}.
Given two fields $\zeta,\eta$ in $\Gamma$ introduce the bounded operators
$A_\kk$ defined by
\be
A_\kk\theta_\kk&=&\langle\eta_\kk|\theta_\kk\rangle\zeta_\kk.
\nnee
Then the diagonal operator $\hat A$ is defined on all of $\Gamma$.
To prove this use that the map $\kk\mapsto \langle\eta_\kk|\theta_\kk\rangle$ is continuous.

\subsection{Integral operators}

The diagonal operators generalize a certain type of diagonal block matrices.
The analogue of non-diagonal block matrices are then integral operators of the
type defined below.

The integral operator $\hat J$ with measurable kernel $J(\kk,\kkp)$ is defined by
\be
[\hat J\zeta]_\kk&=&
\int\upd\kkp\,J(\kk,\kkp)\zeta_\kkp.
\nnee
The domain of definition of $\hat J$ is the subspace of $\Gamma$ consisting of
all $\zeta$ for which
\begin{itemize}
\item $\zeta_\kkp$ is in the domain of $J(\kk,\kkp)$ for all $\kk$ and almost all $\kkp$;
 \item the map $\kkp\mapsto J(\kk,\kkp)\zeta_\kkp$ is integrable for all $\kk$;
 \item the map $\kk\mapsto \int\upd\kkp\,J(\kk,\kkp)\zeta_\kkp$ is continuous.
\end{itemize}
Formally, a diagonal operator $\hat A$ is an integral operator $\hat J$ with kernel
$J(\kk,\kkp)=\delta(\kk-\kkp)A_\kk$. However, this kernel does not satisfy the condition
of integrability.

Given kernels $J(\kk,\kkp)$ and $L(\kk,\kkp)$ the product of the operators $\hat J$ and $\hat L$
involves a convolution of their kernels and can be written as $\hat J\hat L=\widehat{(J*L)}$.
This follows from
\be
[\hat J\hat L\zeta]_\kk
&=&
\int\upd\kkp\,J(\kk,\kkp)[\hat L\zeta]_\kkp\cr
&=&
\int\upd\kkp\,J(\kk,\kkp)\int\upd\kkpp\,L(\kkp,\kkpp)\zeta_\kkpp\cr
&=&
\int\upd\kkpp\,\left(\int\upd\kkp\,J(\kk,\kkp)L(\kkp,\kkpp)\right)\zeta_\kkpp\cr
&=&
[\widehat{(J*L)}\zeta]_\kk
\nnee
with the convolution of kernels $J$ and $L$ defined by
\be
(J*L)(\kk,\kkpp)&=&\int\upd\kkp\,J(\kk,\kkp)L(\kkp,\kkpp).
\nnee

\subsection{Adjoint operators}

The adjoint $\hat A\hc$ of an operator $\hat A$ on $\Gamma$ is
an operator on $\Gammadual$ satisfying 
\be
(\hat A\hc \theta,\zeta)_\kk&=&(\theta,\hat A\zeta)_\kk
\quad\mbox{ for all }\kk\in K, \theta\in\Gammadual, \zeta\in\Gamma.
\nnee
The operator $\hat A$ on $\Gamma$ is said to be symmetric if
$\hat A\hc\zeta\trans=(\hat A\zeta)\trans$ for all $\zeta\in\Gamma$.

\begin{proposition}
\label{prop:lin:adj:exist}
Consider an operator $\hat A$ on $\Gamma$, which is everywhere defined and continuous.
Then there exists a unique adjoint $\hat A\hc$ with domain all of $\Gammadual$.
\end{proposition}

\beginproof
Fix $\theta$ in $\Gammadual$.
Let $\eta\hc_\kk(\zeta)=\theta\hc_\kk(\hat A\zeta)$.
Then the map $\zeta\mapsto \eta\hc_\kk(\zeta)$ is continuous because $\zeta\mapsto \hat A\zeta$ is continuous
by assumption and $\theta\hc_\kk$ belongs to $\Gamma^*$.
In addition is $\kk\mapsto \eta\hc_\kk(\zeta)$ continuous for any $\zeta\in\Gamma$
because $\kk\mapsto \theta\hc_\kk$ is pointwise continuous.
Hence, $\kk\mapsto \eta\hc_\kk$ belongs to $\Gammadual$.

Define the linear operator $\hat A\hc$ by $\hat A\hc \theta=\eta$.
One verifies that
\be
(\hat A\hc \theta,\zeta)_\kk=(\eta,\zeta)_\kk=\overline{\eta_\kk(\zeta)}
=\overline{\zeta_\kk(\hat A\zeta)}=(\zeta,\hat A\zeta)_\kk.
\nnee
This shows that $\hat A\hc$ is an adjoint of $\hat A$.

Assume now that $(\zeta,\theta)_\kk=(\zeta,\eta)_\kk$ for all $\zeta$ and $\kk$.
This means $\theta_\kk(\zeta)=\eta_\kk(\zeta)$ so that the functions $\theta_\kk$ and $\eta_\kk$
coincide for all $\kk$. This implies $\theta=\eta$ and hence uniqueness of the adjoint $\hat A\hc$.

\endproof

Consider a diagonal operator $\hat A$ defined by bounded operators $A_\kk$.
Then $\hat A$ is continuous and everywhere defined. Hence the proposition applies
and the adjoint $\hat A\hc$ is well-defined. In addition one has for all $\theta\in\Gamma$
that $\hat A\hc\theta\trans=\eta\trans$ with the field $\eta$ defined by
$\eta_\kk=A\hc_\kk\theta_\kk$. This implies that
$\hat A\hc$ maps the subspace $\Gamma\trans$ of $\Gammadual$ into itself.

On the other hand, if $\hat J$ is an integral operator with kernel $J_{\kk,\kkp}$,
then one cannot expect that $\hat J\hc$ maps the subspace $\Gamma\trans$ of $\Gammadual$ into itself.
Indeed, one calculates
\be
\left(\hat J\hc\theta\trans,\zeta\right)_\kk
&=&
\left(\theta\trans,\hat J\zeta\right)_\kk\cr
&=&
\left\langle\theta_\kk|[\hat J\zeta]_\kk\right\rangle\cr
&=&
\int\upd\kkp\left\langle\theta_\kk|J_{\kk,\kkp}\zeta_\kkp\right\rangle\cr
&=&
\int\upd\kkp\left\langle J\hc_{\kk,\kkp}\theta_\kk|\zeta_\kkp\right\rangle\cr
&=&
\int\upd\kkp\left(\eta\trans(\kk),\zeta_\kkp\right)_\kkp
\nnee
with $\eta_\kkp(\kk)=J\hc_{\kk,\kkp}\theta_\kk$.
This result is not of the form $(\eta\trans,\zeta)_\kk$.

\subsection{Isometries}

Consider an operator $\hat U$ on $\Gamma$ which conserves field
normalization. Continuity of the map $\hat U$ follows from
\be
||[\hat U\zeta]_\kk||=||\zeta_\kk||\quad\mbox{ for all }\kk\in K.
\nnee
Hence, by Proposition \ref {prop:lin:adj:exist},
$\hat U\hc$ is defined on all of $\Gammadual$. In addition, if $\zeta$ and $\theta$
belong to $\Gamma$ then one has
\be
\left(\hat U\hc(\hat U\theta)\trans,\zeta\right)_\kk
=
\left((\hat U\theta)\trans,\hat U\zeta\right)_\kk
=
\langle [\hat U\theta]_\kk|[\hat U\zeta]_\kk\rangle
=
\langle \theta_\kk|\zeta_\kk\rangle
=
(\theta\trans,\zeta)_\kk.
\nnee
This implies $\hat U\hc(\hat U\theta)\trans=\theta\trans$ for all $\theta\in\Gamma$.

\begin{proposition}
\label{iso:prop2}
 Any strongly continuous map $\kk\mapsto U_\kk$ into the isometries of $\cal H$
 defines a diagonal operator $\hat U$ which conserves field normalization.
\end{proposition}

\beginproof
One has
\be
||U_\kk\zeta_\kk-U_\kkp\zeta_\kkp||
&\le&
||(U_\kk-U_\kkp)\zeta_\kk||+||U_\kkp(\zeta_\kk-\zeta_\kkp)||\cr
&=&
||(U_\kk-U_\kkp)\zeta_\kk||+||\zeta_\kk-\zeta_\kkp||.
\nnee
Hence continuity of the map $\kk\mapsto|| U_\kk\zeta_\kk||$ follows from the strong continuity of
$\kk\mapsto U_\kk$ and continuity of $\kk\mapsto \zeta_\kk$.
This shows that $\hat U\zeta$ belongs to Gamma for all $\zeta$ and therefore that $\hat U$
is defined on all of $\Gamma$. That it conserves field normalization follows immediately.
\endproof

\subsection{Quantum expectations}

The quantum expectation of an operator $\hat A$ on $\Gamma$, given a properly normalized field $\zeta$
belonging to its domain, equals
\be
\langle \hat A\rangle&=&\lu^3\int\upd\kk\,\left(\zeta\trans,\hat A\zeta\right)_\kk,
\nnee
whenever this integral converges. 
The constant length $\lu$ has been added to make the field $\zeta$ dimensionless.
The quantity $\left(\zeta\trans,\hat A\zeta\right)_\kk$
is interpreted as being the quantum expectation of $\hat A$ conditioned on the knowledge of the value of
the wave vector $\kk$. This conditioning is meaningful because the reduction of the representation over
the wave vector is a classical, i.e.~non-quantum aspect of the theory.
A weighing of the integration over $\kk$ may be added
if there is physical evidence for it.

\section{The scalar boson field}
\label{sect:scalbos}

The standard representation of quantum mechanics is said to be irreducible because the
only operators commuting with momentum and position operators $ P$ and $ Q$ are the multiples
of the identity operator.
The reducible representation, used in the present work, is built by integrating irreducible representations.
Following the original work of Marek Czachor and coworkers (See \cite{CM00,CM04,CN06,CW09} and papers cited in these works)
integration over the wave vector $\kk$ is used to decompose the reducible representation
into irreducible ones. This means that for a given wave vector $\kk$ the standard representation
of quantum mechanics in a Hilbert space $\cal H$ is used.
The dependence of the wave vector involves a field of Hilbert spaces $\Gamma$.
It is the linear space which consists of all continuous fields
$\zeta:\,\kk\in\Rotn\mapsto \zeta_\kk\in {\cal H}$.
Note the exclusion of $\kk=0$. It is assumed that the wave vector of a massless boson field
cannot vanish.

\subsection{The irreducible components}

A scalar boson at a given wave vector $\kk$ in $\Rotn$ is described by a quantum harmonic oscillator.

The Hilbert space $\cal H$ equals the space ${\cal L}^2(\Ro,\Co)$ of quadratically integrable
complex functions over the real line.
The momentum operator $P$ and the position operator $Q$ are self-adjoint operators
defined in the usual manner. The annihilation operator
$a$, and its adjoint $a\hc$, are defined by
\be
a&=&\frac{1}{r\sqrt 2} Q+i\frac{r}{\hbar\sqrt 2} P.
\nnee
The positive constant length $r$ is introduced to make the operators $a$, and $a\hc$ dimensionless.
The Hamiltonian $ H$ of the harmonic oscillator can then be written as
\be
 H&=&\hbar\omega a\hc a,
\nnee
with $\omega>0$ the frequency of the oscillator.
Note that the so-called ground state energy is omitted.
In what follows the frequency $\omega$ will depend on a 3-dimensional wave vector $\kk$,
with a  so-called {\em linear dispersion relation}
\be
\omega(\kk)&=&c|\kk|.
\nnee
Here, $c$ is the speed of light. 

The eigenstates of the Hamiltonian $H$ are denoted $|n\rangle$, $n=0,1,2\cdots$.
They can be constructed starting from the ground state $|0\rangle$ by the action of the
creation operator $a\hc$.

\subsection{Coherent states}

Fix a complex number $z$. The following wave function determines a {\em coherent state}
\be
|z\ranglec&=&e^{-\frac 12|z|^2}\sum_{n=0}^\infty \frac{1}{\sqrt{n!}}z^n|n\rangle
\nnee
The sum is convergent and the wave function is normalized to one:
\be
||\,|z\ranglec||=\sqrt{\langlec z|z\ranglec}=1.
\nnee
All coherent states belong to the domain of the annihilation operator $a$
and satisfy
\be
a|z\ranglec&=&z|z\ranglec.
\nnee
They also belong to the domain of the creation operator $a\hc$.
The maps $w\rightarrow |w\ranglec$  and  $|w\ranglec\rightarrow w$ are one-to-one and continuous.

\subsection{Coherent fields}
\label{sect:red:cohfield}

Let be given a continuous complex function $F(\kk)$.
Use it to define the  wave function $|F\ranglec$ of a {\em coherent field} by $[|F\ranglec]_\kk=|F(\kk)\ranglec$.
This coherent field is a properly normalized element of $\Gamma$. 
Clearly is
\be
a[|F\ranglec]_\kk=F(\kk)[|F\ranglec]_\kk\quad\mbox{ for all }\kk\in\Rotn.
\nnee
Coherent fields play an important role further on in the development of the free field theory.

With some abuse of notation the constant field $\kk\mapsto |0\ranglec$
will be denoted $|0\ranglec$ as well as $|0\rangle$.
It is the vacuum state of the free field theory.

The extension $\hat a$ of the annihilation operator $a$ to a diagonal operator on $\Gamma$
satisfies the following property, which can be proved easily.

\begin{proposition}
\label{prop:cohdoma}
Let be given a continuous complex function $F(\kk)$. The coherent field
$|F\ranglec$ belongs to the domain of the diagonal operator $\hat a$
and satisfies $[\hat a|F\ranglec]_\kk=F(\kk)|F(\kk)\ranglec$ for all $\kk$ in $\Rotn$.
If $F(\kk)$ is bounded then $\hat a|F\ranglec$ is bounded in norm.
\end{proposition}

A similar result holds for the creation operator $\hat a\hc$.

\begin{proposition}
\label{prop:cohdomastar}
Let be given a continuous complex function $F(\kk)$. The coherent field
$|F\ranglec$ belongs to the domain of the diagonal operator $\hat a\hc$.
If $F(\kk)$ is bounded then $\hat a\hc|F\ranglec$ is bounded in norm.
\end{proposition}

\subsection{The free field Hamiltonian}

The free-field Hamiltonian $\hat H$ is an unbounded symmetric operator on $\Gamma$.
It is the diagonal operator defined by
\be
[\hat H\zeta]_\kk= H_\kk\zeta_\kk\quad\mbox{ with }\quad
H_k=\hbar c|\kk|a\hc a,
\label{linop:ex:Ham}
\ee
where $a$ is the annihilation operator introduced before.
Its domain of definition is the subspace of $\Gamma$ consisting of all $\zeta$ in $\Gamma$
such that
\begin{itemize}
 \item $\zeta_\kk$ is in the domain of the self-adjoint operator $a\hc a$ for all $\kk$;
 \item $\kk\in\Rotn\mapsto |\kk|a\hc a\zeta_\kk$ is continuous.
\end{itemize}

Physically acceptable free fields necessarily
are superpositions with the vacuum field. Only then it is feasible to obtain a finite value for the
expectation value of the energy operator $H$.

\subsection{The classical wave equation}
\label{sect:scalbos:waveeq}

A large class of solutions of the free wave equation $\square_x \phi=0$ consists
of functions $\phi(x)$ of the form
\be
\phi(x)&=&2\Re\int\upd\kk\,\frac{\lu^{3/2}}{\Nph (\kk)}f(\kk)e^{-ik_\mu x^\mu},
\label{scalar:clas:sol}
\ee
where $f$ is a continuous function of $\Rotn$.

The so-called normalization factor $\Nph (\kk)$ is the usual one
\be
\Nph (\kk)=\sqrt{(2\pi)^32|\kk|\lu},
\label{scalar:class:norm}
\ee
except that the constant $\lu$ is inserted also here to make it dimensionless.
The insertion of this normalization factor leads further on to a satisfactory physical interpretation
of the profile function $f(\kk)$.

The total energy of the classical field $\phi$ is given by
\be
\Hcl
&=&\frac {\hbar c}{2\lu^2}\int_{\Ro^3}\upd\xx\,\left[\left(\frac{\partial\phi}{\partial x^0}\right)^2
+\sum_\alpha\left(\frac{\partial\phi}{\partial x^\alpha}\right)^2
\right].
\label{scalar:classenerg}
\ee
From (\ref {scalar:clas:sol}) one obtains
\be
\Hcl&=&\int_{\Ro^3}\upd\kk \,\hbar c|\kk||f(\kk)|^2.
\label{scalar:clas:classen}
\ee
The interpretation in the context of quantum mechanics is standard.
The factor $|f(\kk)|^2$ is the density of particles with wave vector $\kk$
and corresponding energy $\hbar c|\kk|$.

The particle density $|f(\kk)|^2$ has the dimension of the inverse of a volume in $\kk$-space. Introduce therefore
the dimensionless function
\be
F(\kk)&=&l^{-3/2}f(\kk)
\nnee
and use it to construct the coherent field $|F\ranglec$  (See Appendix \ref {sect:red:cohfield}).
Consider now the Hamiltonian $\hat H$ of the free boson field as given by (\ref {linop:ex:Ham}).
Its irreducible components satisfy
\be
\langlec F(\kk)| H_\kk|F(\kk)\ranglec
&=&
\hbar c|\kk|\langlec F(\kk)|a\hc a|F(\kk)\ranglec\cr
&=&
\hbar c|\kk|\,|F(\kk)|^2\cr
&=&
\lu^{-3}\hbar c|\kk||f(\kk)|^2.
\nnee
This result allows us to write the classical energy (\ref{scalar:clas:classen}) 
in terms of the free-field Hamiltonian $\hat H$ and the coherent field $|F\ranglec$
\be
\Hcl&=&\lu^3\int_{\Ro^3}\upd\kk \,\langle F(\kk)|H_\kk|F(\kk)\ranglec\le +\infty.
\label{scalar:clas:classen2}
\ee

\subsection{Correspondence principle}

Introduce field operators $\hat \phi(x)$, with $x$ in Minkowski space $\Ro^4$, defined by
\be
[\hat\phi(x)\zeta]_\kk=\phi_\kk(x)\zeta_\kk
\nnee
with
\be
\phi_\kk(x)&=&\frac{1}{\Nph (\kk)}\left(e^{-ik_\mu x^\mu}  a +  e^{ik_\mu x^\mu} a\hc\right).
\label{bos:cor:fdop}
\ee
The eigenstates $|n\rangle$, $n=0,1,\cdots$ of the harmonic oscillator belong to the domain
of the r.h.s.~of (\ref {bos:cor:fdop}), as well as all coherent states $|z\rangle$, $z\in\Co$.
It is obvious to define $\phi_\kk$ as the self-adjoint extension of the r.h.s.~of (\ref {bos:cor:fdop}).
The map $\kk\mapsto \phi_\kk(x)$ defines a diagonal operator $\hat \phi(x)$ of $\Gamma$.
It is called the {\em free field operator}.

The free field operators satisfy the commutation relations
\be
[\hat\phi(x),\hat\phi(y)]_-
&=&
\left(\kk\mapsto\frac{1}{(2\pi)^3\lu |\kk|}i\sin(k_\mu(y^\mu-x^\mu)\right).
\nnee
The r.h.s.~of this expression is a bounded diagonal operator which commutes with all
other diagonal operators of $\Gamma$.

Derivatives to all orders of $\hat \phi(x)$ with respect to $x^\mu$ are again diagonal operators.
In particular the free field operators satisfy the operator-valued wave equation
\be
\square_x\hat \phi(x)&=&0.
\nnee

\begin{proposition}
Given any continuous complex function $f$ of $\Rotn$ and the 
corresponding function $F(\kk)=\lu^{-3/2}f(\kk)$,
the coherent field $|F\ranglec$  belongs to the domain of the free field operator $\hat \phi(x)$
for any $x$ in Minkowski space $\Ro^4$.
\end{proposition}

\section{Electromagnetic fields}
\label{section:emf}

The vector potential $A_\mu(x)$ of classical electromagnetism has a so-called gauge freedom.
This means that it is not fully determined by the physical quantities which are the electric
and magnetic forces. It is tradition to fix this freedom by use of the Lorentz gauge.
It has the advantage of leading to a theory which is manifestly Lorentz covariant.
However, it does not eliminate all freedom of choice. The description of free electromagnetic
fields is most convenient in the so-called transverse gauge. It limits the number of degrees
of freedom to two transversely propagating electromagnetic waves. 

\subsection{The classical vector potential}

An electromagnetic wave traveling in direction 3 with electric
component in direction 1 can be described by the vector potential
\be
\Acl(x)&\sim&
\left(
 \begin{array}{c}
 0\\1\\0\\0
 \end{array}\right)\cos(\kph(x^3-x^0)).
\nnee
Here $\kph$ is the wave vector. The index 'ph' is used to label wave vectors of the electromagnetic field.
The electric and magnetic fields can be derived from the vector potential $\Acl(x)$ by
 \be
 \Ecl_\alpha&=&-\frac{\partial \Acl_\alpha}{\partial t}-c\frac{\partial \Acl_0}{\partial x^\alpha},\cr
 \Bcl_\alpha&=&\sum_{\beta,\gamma}\epsilon_{\alpha,\beta,\gamma}\frac{\partial \Acl_\gamma}{\partial x^\beta}.
 \nnee
One then finds
\be
\Ecl_1&\sim& ck\sin(\kph(x^3-x^0)),
\nnee
and $\displaystyle \Bcl_2=-\frac 1c\Ecl_1$ and  $\Ecl_2=\Ecl_3=c\Bcl_1=c\Bcl_3=0$.

Now let $\Xi(\kkph)$ be a rotation matrix which rotates the arbitrary 
 wave vector $\kkph\in\Rotn$ into the positive $z$-direction.
Then an electromagnetic wave with wave vector $\kkph$ is described by the vector potential
with components $\Acl_0(x)=0$ and
\be
\Acl_\alpha(x)&\sim&\Re \Xi_{1,\alpha}(\kkph)e^{-i\kph_\mu x^\mu}.
\nnee
After smearing out with a complex weight function $f(\kkph)$,
and inserting a normalization factor as before (See (\ref {scalar:class:norm})), this becomes
\be
\Acl_\alpha(x)=\Re 
\int\upd\kkph \,\frac{\lambda\lu^{3/2}}{\Nph (\kkph)}f(\kkph)
\Xi_{1,\alpha}(\kkph)e^{-i\kph_\mu x^\mu}.
\nnee
The parameter $\lambda$ could be absorbed into the weight function $f(\kkph)$.
However,  it is kept for dimensional reasons.

The free electromagnetic wave has two possible polarizations.
The second linear polarization is obtained by replacing 
$\Xi_{1,\alpha}(\kkph)$ by $\Xi_{2,\alpha}(\kkph)$ in the previous expression.
In addition, the two polarizations can be combined by adding up the corresponding vector
potentials. The general expression is of the form
\be
\Acl_\alpha(x)= \Re 
\int\upd\kkph \,\frac{\lambda\lu^{3/2}}{\Nph (\kkph)}
\sum_{\beta=1,2}f_\beta(\kkph)
\Xi_{\beta,\alpha}(\kkph)e^{-i\kph_\mu x^\mu}.
\label{em:vp}
\ee

\subsection{Field operators}
\label{em:fieldop}

Because the electromagnetic wave has two polarizations it is obvious to consider a 2-dimensional
quantum harmonic oscillator instead of the single oscillator used in Appendix \ref {sect:scalbos}
on scalar bosons. 

Let $\ah$ and $\av$ be the annihilation operators for a photon with horizontal respectively
vertical polarization.
The free-field Hamiltonian of the quantized electromagnetic field $\hat \Hph$ 
is the diagonal operator defined by
\be
\Hph_\kkph=\hbar c|\kkph|\left(\ahdagger\ah+\avdagger\av\right).
\label{em:ham}
\ee
Field operators $\hat A_\alpha(x)$ are defined by
\be
 A_{\alpha,\kkph}(x)&=&\frac {\lambda}{2\Nph (\kkph)}\varepsilon^{(H)}_\alpha(\kkph)
\left[e^{-i\kph_\mu x^\mu}\ah+e^{i\kph_\mu x^\mu}\ahdagger\right]\cr
& &+\frac {\lambda}{2\Nph (\kkph)}\varepsilon^{(V)}_\alpha(\kkph)
\left[e^{-i\kph_\mu x^\mu}\av+e^{i\kph_\mu x^\mu}\avdagger\right],
\label{em:potop}
\ee
with polarization vectors $\varepsilon^{(H)}_\alpha(\kkph)$ and $\varepsilon^{(V)}_\alpha(\kkph)$
given by two rows of the rotation matrix $\Xi$
\be
\varepsilon^{(H)}_\alpha(\kkph)
=\Xi_{1,\alpha}(\kkph)
\quad\mbox{ and }\quad
\varepsilon^{(V)}_\alpha(\kkph)
=\Xi_{2,\alpha}(\kkph).
\nnee
Note that $\ah$ and $\av$ commute and that
$[\ah,\ahdagger]= \Io$ and $[\av,\avdagger]= \Io$.
This can be used to verify
that the field operators $\hat A_\alpha(x)$ satisfy Heisenberg's equation of motion
\be
i\hbar c\partial_0 \hat A_\alpha(x)&=&\left[\hat A_\alpha(x),\hat\Hph\right]_-.
\nnee

Fix a properly normalized field $\zeta$ in $\Gamma$.
The quantum expectation of the field operators becomes
\be
\Acl_\alpha(x)
&=&
\lu^3 \int\upd\kkph \left(\zeta,| A_\alpha(x)\zeta\right)_{\kkph}\cr
&=&
\lu^3 \Re\int\upd\kkph \frac{\lambda}{\Nph (\kkph)}e^{-i\kph_\mu x^\mu}
\cr & &\times
\left[
\varepsilon^{(H)}_{\alpha}(\kkph)
\langle\zeta_{\kkph}|\ah\zeta_{\kkph}\rangle
+\varepsilon^{(V)}_{\alpha}(\kkph)
\langle\zeta_{\kkph}|\av\zeta_{\kkph}\rangle\right].
\label{photon:classfield}
\ee
This is of the form (\ref {em:vp}) with
\be
f_1(\kkph)=\lu^{3/2}\langle\zeta_{\kkph}|\ah\zeta_{\kkph}\rangle
\quad\mbox{ and }\quad
f_2(\kkph)=\lu^{3/2}\langle\zeta_{\kkph}|\av\zeta_{\kkph}\rangle.
\nnee

Operator-valued electric and magnetic fields are defined by
 \be
 \hat E_\alpha&=&-c\partial_0 \hat A_\alpha,\cr
 \hat B_\alpha&=&\sum_{\beta,\gamma}\epsilon_{\alpha,\beta,\gamma}\frac{\partial \,}{\partial\xx_\beta}
\hat A_\gamma.
 \nnee
Gauss's law in absence of charges is satisfied. Indeed, the divergence of the electric field operators vanishes,
as follows from
\be
\sum_\alpha\partial_\alpha E_{\alpha,\kkph}
&=&\frac 1{2\Nph (\kkph)} \lambda c|\kkph|\left(\sum_\alpha \kph_\alpha \varepsilon^{(H)}_\alpha(\kkph)\right)
\left[ e^{-i\kph_\mu x^\mu}\ah+e^{i\kph_\mu x^\mu}\ahdagger\right]
\cr
& &+\frac 1{2\Nph (\kkph)} \lambda c|\kkph|\left(\sum_\alpha \kph_\alpha \varepsilon^{(V)}_\alpha(\kkph)\right)
\left[ e^{-i\kph_\mu x^\mu}\av+e^{i\kph_\mu x^\mu}\avdagger\right]
\cr
&=&0,
\label{photon:gauss}
\ee
because
\be
\sum_\alpha \kph_\alpha \varepsilon^{(H)}_\alpha(\kkph)
=\left(\Xi(\kkph)\kkph\right)_1
=|\kkph|(\eev_3)_1
\label{em:hororth}
\ee
vanishes, as well as a similar expression for the vertical polarization.

Finally let us calculate the commutation relations
\be
\left[A_{\alpha,\kkph}(x),A_{\beta,\kkph}(y)\right]_-
&=&-\frac i{(2\pi)^34|\kkph|\lu}\lambda^2\sin( \kph_\mu(x-y)^\mu)\cr
& &\times
\left(
\varepsilon^{(H)}_\alpha(\kkph)\varepsilon^{(H)}_\beta(\kkph)
+\varepsilon^{(V}_\alpha(\kkph)\varepsilon^{(V)}_\beta(\kkph)\right).
\nnee
These commutation relations differ from the standard ones
in the first place because the integration over the $\kkph$ vector,
found in the standard theory, is missing. 

\subsection{Coherent fields}

A pair of complex numbers $z,w$ determines a coherent state of the two-dimensional quantum harmonic oscillator.
It satisfies $\ah|z,w\ranglec=z|z,w\ranglec$ and $\av|z,w\ranglec=w|z,w\ranglec$.
Given two complex continuous functions $f_1(\kkph)$, $f_2(\kkph)$ a properly normalized 
field $\zeta$ in $\Gamma$ is defined by
\be
\zeta_\kkph&=&|F_1(\kkph),F_2(\kkph)\ranglec,\qquad \kkph\in\Rotn,
\nnee
with $F_i(\kkph)=\lu^{-3/2}f_i(\kkph)$, $i=1,2$.
This field $\zeta$ describes a coherent electromagnetic field.
It belongs to the domain of the free Hamiltonian $\hat \Hph$.
The quantum expectation of the latter equals
\be
\Hqu
&=&
\lu^3\int\upd\kk\,(\zeta,\hat H\zeta)_\kk\cr
&=&
\int\upd\kk\,\hbar c|\kkph|\left(|f_1(\kkph)|^2+|f_2(\kkph)|^2\right)
\le+\infty.
\nnee
The interpretation is obvious: $|f_1(\kkph)|^2$ and $|f_2(\kkph)|^2$
are the expected densities for horizontally, respectively vertically
poarized photons with kinetic energy $\hbar c|\kkph|$.

\subsection{Single photon states}
\label{sect:em:single}

An important example of an incoherent field is the electromagnetic field produced by a
single photon. In the present formalism this field requires a wave vector-dependent
superposition of the single photon wave function, say $|1,0\rangle$ for a horizontally
polarized photon, with the ground state $|0,0\rangle$ of the two-dimensional harmonic
oscillator. This superposition can be written as
\be
\zeta_\kkph=\sqrt{\rho(\kkph)}e^{i\phi(\kkph)}|1,0\rangle+\sqrt{1-\rho(\kkph)}|0,0\rangle.
\nnee
The energy of the electromagnetic wave equals
\be
\Hqu&=&\hbar c\int\upd\kkph\,|\kkph|\rho(\kkph).
\nnee
The wave vector distribution $|\kkph|\rho(\kkph)$ must be integrable to keep the total energy finite.
In particular, $\rho(\kkph)$  cannot be taken constant.
Therefore, the superposition of the one-photon wave function and the ground state wave function is a necessity.

The quantum expectation of the vector potential evaluates to
\be
\Acl_\alpha(x)
&=&
\lambda\int\upd\kkph\, \frac{\lu^{5/2}}{\Nph (\kkph)}\sqrt{\rho(\kkph)(1-\rho(\kkph))}
\varepsilon^{(H)}_{\alpha}(\kkph)
\Re e^{i\phi(\kkph)}e^{-i\kph_\mu x^\mu}.\cr
& &
\nnee
Note that the contribution to the classical electromagnetic field comes from the region where
the overlap with the ground state is neither 0 nor 1.

The one-photon field discussed above is linearly polarized.
An example of circularly polarized one-photon field is obtained by choosing
\be
\zeta_\kkph=
\sqrt{\rho(\kkph)}e^{i\phi(\kkph)}
\frac{1}{\sqrt 2}(|1,0\rangle\pm i|0,1\rangle)
+\sqrt{1-\rho(\kkph)}|0,0\rangle.
\nnee
The spin is given by
$\displaystyle
S_2=\pm \lu^{3}\int\upd\kkph\,\rho(\kkph)
$.
Note that
$\displaystyle
\frac{1}{\sqrt 2}(|1,0\rangle\pm i|0,1\rangle)
$
is a wave function of the 2-dimensional harmonic oscillator, just like $|1,0\rangle$
or $|0,1\rangle$. Both linearly and circularly polarized one-photon fields exist
in the present theory.

\section{Scalar fermions}

\subsection{The Klein-Gordon equation}

This section concerns the quantum field description of fermions with
a rest mass $m>0$. The appropriate wave equation is the Klein-Gordon equation
\be
(\square +\mstar^2)\phi(x)=0
\quad\mbox{with }\mstar=\frac{mc}{\hbar}.
\label{fermion:kg}
\ee
For $m=0$ it reduces to the d'Alembert equation $\square\phi=0$,
discussed in Appendix \ref {sect:scalbos:waveeq}.
Propagating wave solutions are of the same form as in Appendix \ref {sect:scalbos:waveeq}
\be
\phi(x)
&=&
2\Re \int_{\Ro^3}\upd\kk \,\frac{\lu^{3/2}}{\Nel(\kk)}
f(\kk)e^{-ik_\mu x^\mu},
\label{weq:exp}
\ee
but with a dispersion relation given by the positive square root
\be
k^0=\omega(\kk)/c
\quad\mbox{with}\quad \omega(\kk)=c\sqrt{\mstar^2+|\kk|^2},
\nnee
and a corresponding normalization
\be
\Nel(\kk)=\sqrt{(2\pi)^32\lu \omega(\kk)/c}.
\nnee
The constant $\lu$ is inserted in (\ref {weq:exp}) for dimensional reasons. 
It makes $|f(\kk)|^2$ into a density.

\subsection{Larmor precession}
\label{fermion:larmor}

We use the harmonic oscillator in the description of bosons because it
exhibits periodic motion.
An alternative model exhibiting periodic motion is that of Larmor precession.
It involves the Pauli matrices $\sigma_\alpha$, $\alpha=1,2,3$.
The time evolution is
\be
 \sigma_1(t)&=& \sigma_1\cos(\omega t)+ \sigma_2\sin(\omega t),\\
 \sigma_2(t)&=& \sigma_2\cos(\omega t)- \sigma_1\sin(\omega t),\\
 \sigma_3(t)&=& \sigma_3.
\nnee
The Hamiltonian reads
\be
 H=-\frac{1}{2}\hbar\omega  \sigma_3.
\label{fermion:larmorham}
\ee

\subsection{Fermionic state space}

Let $\Gamma_2$ denote the linear space of continuous fields $\zeta:\,\kk\in\Ro^3\mapsto \zeta_\kk\in \Co^2$.
Like in the case of bosonic fields it is a locally convex Hausdorff space.
However, because the Hilbert space $\Co^2$ is finite-dimensional it is also a Banach space.
An element $\zeta$ of $\Gamma_2$ is said to be properly normalized if $||\zeta_\kk||=1$ for all $\kk$.
States of the fermionic quantum field theory are represented by properly normalized fields.

Note that any properly normalize field $\zeta$ of $\Gamma_2$ can be written into the form
\be
\zeta_\kk=\left(\begin{array}{c}
               \sqrt{1-\rho(\kk)}e^{i\chi(\kk)}\\
               \sqrt{\rho(\kk)}e^{i\xi(\kk)}
               \end{array}\right),
\label{fermion:wf}
\ee
where $\rho$, $\chi$ and $\xi$ are real-valued functions of $\kk\in\Ro^3$.
By adopting this way of writing one tacitly assumes that  $(1,0)^{\rm T}$ means absence of the fermion,
while $(0,1)^{\rm T}$ means presence of the fermion. Note the analogy with single photon states as described in
Appendix \ref {sect:em:single}.
With this interpretation $\rho(\kk)$ becomes the density of the fermion field with wave vector $\kk$.

The Hamiltonian (\ref{fermion:larmorham}) of Larmor precession defines a diagonal operator $\hat\Hel$
by $[\hat \Hel\zeta]_\kk=\Hel_\kk\zeta_\kk$ with
\be
\Hel_\kk=\frac{1}{2}\hbar\omega(\kk) (\Io-\sigma_3).
\label{fermion:hamplus}
\ee
A constant matrix has been added to make the Hamiltonian non-negative.
This does not change the dynamics of the Larmor precession.
The domain of definition of $\hat\Hel$ is all of $\Gamma_2$.

With the help of (\ref {fermion:wf}) the quantum expectation of the Hamiltonian becomes
\be
\langle\hat\Hel\rangle
&=&\lu^3\int\upd\kk\,\left(\zeta,\hat H\zeta\right)_\kk\cr
&=&\lu^3\int\upd\kk\,\hbar\omega(\kk)\rho(\kk).
\label{fermion:quantenerg}
\ee
This reveals that $\rho(\kk)$ is a distribution of quantum particles with dispersion relation $\omega(\kk)$.
It is restricted by the condition that $0\le\rho(\kk)\le 1$ for all $\kk$.
Because the energy must remain finite the distribution $\rho(\kk)$ should go to 0
fast enough for large values of the wave vector $|\kk|$.

\subsection{Field operator}

Introduce now the field operator $\hat\phi(x)$ defined by $[\hat\phi(x)\zeta]_\kk=\phi_\kk\zeta_\kk$ with
\be
\phi_\kk(x)=\frac{1}{\Nel(\kk)}\left[\sigma_+(t)e^{i\kk\cdot\xx}+\sigma_-(t)e^{- i\kk\cdot\xx}\right].
\label{fermion:fieldop}
\ee
It is tradition to decompose this field operator into so-called positive-frequency and negative-frequency
operators
\be
\hat\phi(x)&=&\hat\phi^{(+)}(x)+\hat\phi^{(-)}(x), \quad\mbox{ with }\cr
\phi^{(+)}_\kk(x)&=&\frac{1}{\Nel(\kk)}\sigma_+(t)e^{i\kk\cdot\xx}=\frac{1}{\Nel(\kk)}\sigma_+e^{-ik_\nu x^\nu},\cr
\phi^{(-)}_\kk(x)&=&\frac{1}{\Nel(\kk)}\sigma_-(t)e^{- i\kk\cdot\xx}=\frac{1}{\Nel(\kk)}\sigma_-e^{ik_\nu x^\nu}.
\nnee
They satisfy the anti-commutation relations
\be
\hat\phi^{(+)}(x)\hat\phi^{(+)}(y)&=&0,\cr
\{\hat\phi^{(+)}(x),\hat\phi^{(-)}(y)\}_+
&=&
\left(\kk\mapsto \frac{c}{2(2\pi)^3\lu\omega(\kk)} e^{-ik_\mu (x-y)^\mu}\right).
\label{fermion:car}
\ee
These anti-commutation relations are non-canonical.
Note that
\be
\hat\phi^{(-)}(x)&=&\left(\hat\phi^{(+)}(x)\right)\hc.
\nnee

The classical field $\phicl(x)$ corresponding with the field operator $\hat\phi(x)$ equals
\be
\phicl(x)
&=&\lu^3\int\upd\kk \,\left(\zeta,\hat\phi(x)\zeta\right)_\kk\cr
&=&\lu^3\int\upd\kk \,\langle\zeta_\kk|\frac{1}{N(k)}
\left[\sigma_+e^{-ik_\mu x^\nu}+\sigma_-e^{ik_\mu x^\nu}\right]
\zeta_\kk\rangle\cr
&=& \int\upd\kk \,\frac{\lu^{3}}{\Nel(\kk)}
\sqrt{\rho(\kk)(1-\rho(\kk))}2\Re e^{-i(\chi(\kk)-\xi(\kk))}e^{-ik_\mu x^\mu}.
\label{fermion:classfield}
\ee
The integral converges provided that the density $\rho(\kk)$ tends fast enough
 either to 0 or to 1 for large values of $|\kk|$.
The expression (\ref {fermion:classfield}) is
of the form (\ref {weq:exp}) with
\be
f(\kk)&=&\lu^{3/2}\sqrt{\rho(\kk)(1-\rho(\kk))}e^{-i(\chi(\kk)-\xi(\kk))}.
\nnee

\section{The free Dirac equation}

\subsection{The algebra of creation and annihilation operators}

The electron wave is fermionic. It has two polarizations,
which are related to the spin of the electron.
In addition, the electron has an anti-particle, which is the positron.
This means that the electron field has 4 internal degrees of freedom
and that we need 4 copies of the spin matrices $\sigma_\pm$
instead of the single copy introduced in Appendix \ref {fermion:larmor}.
The corresponding matrices are denoted $\sigma^{(\pm)}_s$, with the index $s$
running from 1 to 4.
They satisfy the anti-commutation relations
\be
\left\{\sigma^{(+)}_s,\sigma^{(+)}_t\right\}_+&=&0,\cr
\left\{\sigma^{(+)}_s,\sigma^{(-)}_t\right\}_+&=&\delta_{s,t}.
\nnee
The hermitian conjugate of $\sigma^{(+)}_s$ is $\sigma^{(-)}_s$.
Together they generate an algebra known as a Clifford algebra.
An explicit representation of the operators as 16-by-16 
matrices is easily constructed (See for instance Section 3-9 of \cite{JR76}).
However, it is not needed in the sequel.

Basis vectors of the 16-dimensional Hilbert space ${\cal H}_{16}$
are specified by subsets $\Lambda\subset\{1,2,3,4\}$
and are given by
\be
|\Lambda\rangle&=&
[\sigma_{4}^{(-)}]^{\Io_{4\in\Lambda}}
[\sigma_{3}^{(-)}]^{\Io_{3\in\Lambda}}
[\sigma_{2}^{(-)}]^{\Io_{2\in\Lambda}}
[\sigma_{1}^{(-)}]^{\Io_{1\in\Lambda}}|\emptyset\rangle.
\nnee
For instance, if $\Lambda=\{1,3\}$ then
$|\{1,3\}\rangle=\sigma_{3}^{(-)}\sigma_{1}^{(-)}|\emptyset\rangle$.

The field operators $\hat\phi_s(x)$ are diagonal operators on $\Gamma_2$ defined by matrices
$\phi_{s,\kk}(x)$. The latter can be written in the form (\ref {fermion:fieldop}).
They satisfy the anti-commutation relations 
\be
\{\phi_{s,\kk}^{(+)}(x),\phi_{t,\kkp}^{(+)}(y)\}_+&=&0
\quad\mbox{ and }\cr
\{\phi_{s,\kk}^{(+)}(x),\phi_{t,\kkp}^{(-)}(y)\}_+&=&
\frac{c}{2(2\pi)^3\lu\omega(\kk)}
\delta_{s,t}e^{-ik_\mu x^\mu}e^{ik'_\mu y^\mu}.
\label{electron:car}
\ee
Note that (\ref {fermion:fieldop}) implies that
\be
\phi_{s,\kk}^{(+)}(x)&=&\frac{1}{\Nel(\kk)}e^{-ik_\mu x^\mu}\sigma^{(+)}_{s}.
\nnee

A familiar notation for these operators, evaluated at $x=0$, is
\be
b_{\uparrow}=\sigma^{(+)}_{1},\quad
b_{\downarrow}=\sigma^{(+)}_{2},\quad
d_{\downarrow}=\sigma^{(+)}_{3},\quad
d_{\uparrow}=\sigma^{(+)}_{4}.
\nnee
This alternative notation is not used here.

\subsection{The Hamiltonian}

The Hamiltonian of the electron field $\hat\Hel$ is the sum of 4 copies of the scalar
Hamiltonian (\ref {fermion:hamplus}). It is defined by $[\hat\Hel\zeta]_\kk=\Hel_\kk\zeta_\kk$
with
\be
\Hel_\kk=\frac{1}{2}\hbar\omega(\kk)\sum_{s=1}^4 (\Io-\sigma^{3}_s).
\label{fermion:dirac:ham}
\ee
Note that the Hamiltonian is positive.
It is tradition to assign negative energies to positrons and positive energies to electrons.
This tradition is not followed here because it does not make sense.
It is a remainder of Dirac's interpretation of positrons as holes in a sea of electrons.
The alternative treatment assigns the vacuum state to one of the eigenstates of $\sigma_3$
instead of assigning a particle/anti-particle pair to the two eigenstates.
The dimension of the Hilbert space goes up from 4 (the number of components of a Dirac spinor)
to 16. This is meaningful because the Dirac equation considered here is an equation for
field operators and not the original one which holds for classical field spinors
(See (\ref {electron:diracclass}) below).

Number operators $N_s$ are defined by
\be
N_s &=&\sigma^{(-)}_{s}\sigma^{(+)}_{s}
\quad s=1,2,3,4.
\label{electron:numop}
\ee
They appear in the Hamiltonian
\be
\Hel_\kk&=&\hbar\omega(\kk)\sum_{s=1}^4 N_s.
\label{electron:ham}
\ee
The field operators $\hat\phi_{s}^{(+)}(x)$ satisfy Heisenberg's equations of motion
\be
i\hbar c\partial_0\hat\phi_{s}^{(+)}(x)
&=&\left[\hat\phi_{s}^{(+)}(x), \hat \Hel\right]_-.
\nnee

In principle, this is all that is needed for a description of 
electron/positron fields. However,
the notion of electric current is needed for a description of
the interactions between the electromagnetic field and the
electron/positron field. 
The derivation below is far from trivial and follows
the approach initiated by Dirac. Let us start by introducing
Dirac's equation for quantum field operators.

\subsection{Dirac's equation}
\label{sect:dirac}

Introduce the gamma matrices. In the standard representation they read
\be
\gamma_0=\left(\begin{array}{lr}
                \Io &0\\0 &-\Io
               \end{array}\right)
\quad\mbox{and}\quad
\gamma_\alpha=\left(\begin{array}{lr}
                0 &-\sigma_\alpha\\\sigma_\alpha &0
               \end{array}\right).
               \label{fermion:gammadef}
\nnee
Next introduce auxiliary field operators $\hat\psi_r$, $r=1,2,3,4$.
They are called the {\em Dirac field operators} and are defined by
\be
\psi_{r,\kk}(x)
&=&
\sqrt{2\lu k_0}
\left[
\sum_{s=1,2}u_r^{(s)}(\kk)\phi_{s,\kk}^{(+)}(x)
+
\sum_{t=3,4}v_r^{(t)}(\kk)\phi_{t,\kk}^{(-)}(x)
\right]\cr
&=&\frac{1}{\sqrt{(2\pi)^3}}
\left[
\sum_{s=1,2}u_r^{(s)}(\kk)e^{-ik_\mu x^\mu}\sigma^{(+)}_{s}
+
\sum_{t=3,4}v_r^{(t)}(\kk)
e^{ik_\mu x^\mu}\sigma^{(-)}_{t}\right].\cr
& &
\label{electron:defpsi}
\ee
The vectors $u^{(1)},u^{(2)},v^{(3)},v^{(4)}$ are the analogues of the polarization vectors of the photon.
They are partly fixed by the requirement that the vector
with components $\hat\psi_r$ satisfies Dirac's equation
\be
i\gamma^\mu\partial_\mu \hat\psi(x)=\mstar \hat\psi(x).
\label{electron:dirac}
\ee
Indeed, using
\be
\partial_\mu \phi_{s,\kk}^{(\pm)}&=&\mp ik_\mu \phi_{s,\kk}^{(\pm)}
\nnee
one finds that a sufficient condition for (\ref {electron:dirac}) to hold is
\be
\gamma^\mu k_\mu u^{(s)}=\mstar u^{(s)}
\quad\mbox{and}\quad
\gamma^\mu k_\mu v^{(t)}=-\mstar v^{(t)}.
\nnee
Each of these two equations has two independent solutions.
They can be chosen to satisfy the orthogonality relations
\be
\sum_r\overline{u^{(s)}_r}(\kk)u^{(s')}_r(\kk)&=&\delta_{s,s'},\cr
\sum_r\overline{v^{(t)}_r}(\kk)v^{(t')}_r(\kk)&=&\delta_{t,t'},\cr
\sum_r\overline{u^{(s)}_r}(\kk)v^{(t)}_r(-\kk)&=&0.
\label {electron:ortho}
\ee

An electron/positron field is now determined by a properly normalized field $\zeta$ of the form
\be
\zeta_\kk&=&\sum_{\Lambda\subset\{1,2,3,4\}\}}z^{\Lambda}_{\kk}|\Lambda\rangle,
\nnee
with complex coefficients $z^{\Lambda}_{\kk}$ satisfying
\be
\sum_{\Lambda\subset\{1,2,3,4\}}|z^{\Lambda}_{\kk}|^2&=&1,
\qquad\mbox{ for all }\kk.
\nnee
It defines a Dirac spinor containing classical fields by
\be
{\phicl}_r(x)
&=&\lu^3\int\upd\kk\,\langle\zeta_\kk|\psi_{r,\kk}\zeta_\kk\rangle,
\label{electron:classfield}
\ee
whenever the integral converges.
This Dirac spinor $\phicl(x)$ with 4 components satisfies the Dirac equation
\be
i\gamma^\mu\partial_\mu \phicl(x)=\mstar\phicl(x).
\label{electron:diracclass}
\ee

Finally note that each of the Dirac field operators, as constructed here, is not only
a solution of the Klein-Gordon equation
\be
[\square+\mstar]\psi_{r,\kk}=0
\label{electron:kgeq}
\ee
but also of the partial equations
\be
\left[c^2\partial_0^2+[\hbar\omega(\kk)]^2\right]\psi_{r,\kk}(x)&=&0,\cr
\left[\Delta+|\kk|^2\right]\psi_{r,\kk}(x)&=&0.
\nnee
A Lorentz transformation can mix up these two equations.

\subsection{Charge conjugation}

The charge conjugation matrix $C$ is defined by 
\be
C\gamma^\mu C^{-1}=-(\gamma^\mu)^{\rm T}.
\nnee
Using the standard representation of the gamma matrices it equals
$C=i\gamma^2\gamma^0$ (See for instance Section 10.3.2 of \cite{GR96}).
In explicit form is
\be
C&=&\left(\begin{array}{lccr}
           0 &0 &0 &-1\\
           0 &0 &1 &0\\
           0 &-1 &0 &0\\
           1 &0 &0 &0
          \end{array}\right).
\nnee
The main properties of the matrix $C$ are
\begin{itemize}
 \item $C^{-1}=C\hc=C^{\rm T}=-C$;
 \item $\sum_{r'}C_{r,r'}\overline {u_{r'}^{(1)}(\kk)}=v^{(4)}_r(-\kk)$;
 \item $\sum_{r'}C_{r,r'}\overline {u_{r'}^{(2)}(\kk)}=v^{(3)}_r(-\kk)$.
\end{itemize}

\bigskip

The charge conjugation operator $\Cc$ is a linear operator on ${\cal H}_{16}$
with the properties that $\Cc^{-1}=\Cc\hc=-\Cc$ and 
\be
\Cc \psi_{r,\kk}(x)\Cc^{-1}&=&-\sum_{r'}C_{r,r'}\psicha_{r',\kk}(x),\cr
\Cc \psicha_{r,\kk}(x)\Cc^{-1}&=&\sum_{r'}C_{r,r'}\psi_{r',\kk}(x).
\label{ferm:charge:mr}
\ee

\section{The Dirac current}

\subsection{Two-point correlations}

Fix a properly normalized electron field $\zeta$.
A two-point correlation function for the Dirac field operators $\hat\psi_r(x)$ 
is defined by 
\be
\Ga_{r',r}(x,x')
&=&
\lu^3\int\upd\kk \,\int\upd\kkp \,
\langle\zeta_{\kk}|\psicha_{r,\kk}(x)\psi_{r',\kkp}(x')\zeta_{\kkp}\rangle,\cr
& &
\label{fermion:twopointmod}
\ee
whenever the integrals converge.
Note the order of the indices $r,r'$.
A short calculation using Dirac's equation shows that
the vector $\pc(x)$  with 4 components
\be
\pc^\mu(x)=\Tr \gamma^\mu \Ga(x,x)
\nnee
satisfies the continuity equation
\be
0&=&\frac{\partial\,}{\partial {x}^\mu}\Tr \gamma^\mu \Ga(x,x).
\label{fermion:cont}
\ee

The vector $\pc (x)$, introduced above, describes a current,
which however is not yet the electric current.
The components of $\pc (x)$ are real numbers.
Indeed, using $(\gamma^\mu)\hc\gamma^0=\gamma^0\gamma^\mu$ 
one verifies that
\be
\overline{\pc ^\mu(x)}&=&
\overline{\Tr \gamma^\mu \Ga(x,x)}\cr
&=&\Tr{\Ga}\hc(x,x)\gamma^0\gamma^\mu\gamma^0\cr
&=&\sum_{r,r'}
\overline{
\langle\zeta_{\kk}|[\psi_{r,\kk}(x)]\hc\psi_{r',\kkp}(x)\zeta_{\kkp}\rangle}
\gamma^0_{r,r}[\gamma^0\gamma^\mu\gamma^0]_{r',r}
\cr
&=&\sum_{r,r'}
\lu^3\int\upd\kk \,\int\upd\kkp \,
\langle\zeta_{\kkp}|[\psi_{r',\kkp}(x)]\hc\psi_{r,\kk}(x)\zeta_{\kk}\rangle
[\gamma^0\gamma^\mu]_{r',r}
\cr
&=&\sum_{r,r'}\Ga_{r',r}(x,x)\gamma^\mu_{r,r'}\cr
&=&\Tr \Ga(x,x)\gamma^\mu\cr
&=&\pc ^\mu(x).
\nnee

\subsection{The electric current}

The electric current operators $\hat J^\mu(x)$ are  integral operators defined by
the symmetric kernels
\be
J^\mu_{\kk,\kkp}(x)
&=&
\frac 12\uc c\left(
\Pc^\mu_{\kk,\kkp}(x)-\Cc\Pc^\mu_{\kk,\kkp}(x)\Cc^{-1}\right).
\label{electron:jdef}
\ee
Here, $\uc $ is a unit of charge.
The domain of definition of $\hat J^\mu(x)$ consists of the fields $\zeta\in\Gamma_2$ 
for which the integrals
\be
\int\upd\kkp\,J^\mu_{\kk,\kkp}(x)\zeta_\kkp
\nnee
are absolutely convergent.
Because $\hat\Pc$ satisfies the continuity equation also $\hat J$ does.
One can show that
\be
J^\mu_{\kk,\kkp}(x)
&=&\frac {1}2\uc c
\sum_{r,r'}\gamma^\mu_{r,r'}\psicha_{r,\kk}(x)\psi_{r',\kkp}(x)
-\frac {1}2\uc c\sum_{r,r'}\gamma^\mu_{r',r}\psi_{r,\kk}(x)\psicha_{r',\kkp}(x).\cr
& &
\label{appE:Jexpr}
\ee
This is a well-known expression for the Dirac current, adapted to the present context.

The total charge $\hat Q$ is the diagonal operator satisfying
\be
\frac 1c\int\upd \xx \,J^0_{\kk,\kkp}(x)=\delta(\kk-\kkp)Q_\kk.
\label{electron:totcharge}
\ee

One finds
\be
\frac 1c\int\upd \xx \, J^0_{\kk,\kkp}(x)
&=&\uc\delta(\kk-\kkp)\left(N_1+ N_2- N_3-N_4\right).
\nnee
This implies
\be
\hat Q&=&\uc\left(N_1+ N_2- N_3-N_4\right).
\nnee
The obvious interpretation is that the components 1 and 2 of the field describe an electron
with charge $\uc $, and that components 3 and 4 describe a positron with charge $-\uc $.

\section{Interaction of Photons and Electrons}

\subsection{The state space}

The reducible representations of the free electromagnetic field
and the free electron/positron field each have their own
wave vector used to label the irreducible components.
They are denoted $\kkph$, respectively $\kk$. A field $\zeta$
of the interacting system associates with each pair $\kkph\not=0$, $\kk$
of wave vectors a wave function $\zeta_{\kkph,\kk}$ in the product Hilbert space
$\Hcalem\times{\cal H}_{16}$, where $\Hcalem$ is 
the Hilbert space of a two-dimensional harmonic oscillator, and ${\cal H}_{16}$
is the 16-dimensional Hilbert space representing the possible states of an 
electron/positron field.
Basis vectors in the product Hilbert space are denoted
$|m,n\rangle\times|\Lambda\rangle\equiv |m,n,\Lambda\rangle$, where $m,n$ count photons
and $\Lambda$ is a subset of $\{1,2,3,4\}$.
The space of continuous fields of the form
\be
(\kkph,\kk)\in\Rotn\times\Ro^3\mapsto \zeta_{\kkph,\kk}\in \Hcalem\times{\cal H}_{16}.
\nnee
is denoted $\Gammaphel$.

\subsection{The interaction Hamiltonian}

The Hamiltonian $\hat H$ is of the usual form
\be
\hat H&=&\hat\Hph+\hat\Hel+\hatHI.
\label{schroed:fullham}
\ee
The kinetic energy of the photon field is given by (\ref {em:ham})
\be
\Hph_\kkph&=&\hbar c|\kkph|\left(\ah^\dagger\ah+\av^\dagger\av\right),
\nnee
that of the electron/positron field by (\ref {fermion:dirac:ham}, \ref {electron:ham})
\be
\Hel_\kk=\frac{1}{2}\hbar\omega(\kk)\sum_{s=1}^4 N_s,
\nnee
where $N_s$ is the number operator indicating the presence of a particle of type $s$
(electron or positron, spin up or down).
The interaction term involves the Dirac current $\hat J^\mu(x)$
and the electromagnetic potential operators $\hat A_\mu(x)$.
The obvious definition is
\be
\hatHI(x^0)&=&\int_{\Ro^3}\upd \xx\,
\hat A_\mu(x)\hat J^\mu(x).
\label{int:int:int}
\nnee
For $x^0=0$ this defines the interaction Hamiltonian in the Schr\"odinger picture.

\subsection{Gauge transformations}
\label{app:gauge}

The charge operator $\hat Q$ commutes with $\hat \Hph$, $\hat \Hel$ and $\hat \HI$ and hence with
the full Hamiltonian $\hat H$. The one-parameter group $\Lambda\in\Ro\mapsto\exp(i\Lambda \hat Q)$
is a global symmetry of electrodynamics and corresponds with the U(1) gauge group
of the Standard Model. This raises the question whether this symmetry group can be extended
to include local symmetries. Local means here local in the space of wave vectors.

Given a smooth function $\Lambda(\kkph,\kk)$ introduce the diagonal operator
$\hat U_\Lambda$ defined by
\be
[\hat U_\Lambda\zeta]_{\kkph,\kk}&=&e^{i\Lambda(\kkph,\kk) Q}\zeta_{\kkph,\kk}.
\nnee

\begin{proposition}
Assume that any of the 4 possibilities $\kk'=\pm\kk\pm\kkph$
implies that $\Lambda(\kkph,\kk')=\Lambda(\kkph,\kk)$.
Then $\hat U_\Lambda$ commutes with the Hamiltonian $\hat H$.
\end{proposition}

\beginproof

It clearly commutes with $\hat \Hph$ and $\hat \Hel$. 
Because $\hat Q$ commutes with $\hat J^\alpha$ one has
\be
[\hat U_\Lambda^{-1}\hatHI \hat U_\Lambda\zeta]_{\kkph,\kk}
&=&
\int_{\Ro^3}\upd \xx\,
\sum_\alpha A_{\alpha,\kkph}(0,\xx)\cr
& &\times \lu^3\int\upd\kk'\, 
e^{i[\Lambda(\kkph,\kk')-\Lambda(\kkph,\kk)] Q}
J^\alpha_{\kk,\kk'}(0,\xx)\zeta_{\kkph,\kk'}.
\nnee
The $\xx$-dependence of $A_{\alpha,\kkph}(0,\xx)$ involves factors $e^{i\kkph\cdot\xx}$.
The $\xx$-dependence of $J^\alpha_{\kk,\kk'}(0,\xx)$ involves factors $e^{i(\pm\kk\pm\kkp)}$.
Hence the integration over $\xx$ produces Dirac delta functions $\delta(\pm\kkph\pm\kk\pm\kkp)$.
By assumption these restrictions on the wave vectors imply that
$\Lambda(\kkph,\kk')=\Lambda(\kkph,\kk)$.
Therefore the factor $e^{i[\Lambda(\kkph,\kk')-\Lambda(\kkph,\kk)] Q}$ may be omitted
in the above expression. The result is that
\be
[\hat U_\Lambda^{-1}\hatHI \hat U_\Lambda\zeta]_{\kkph,\kk}
&=&
[\hatHI \zeta]_{\kkph,\kk}.
\nnee
This implies that $\hatHI$ commutes with $\hat U_\Lambda$.

\endproof

Let $\kk=\kk_\parallel+\kk_\perp$ be the decomposition of $\kk$
into a part parallel to $\kkph$ and an orthogonal part.
With this notation, the condition of the proposition is satisfied when $\Lambda(\kkph,\kk)$ is a function of
$\kkph$ and $|\kk_\perp|$ only. It is easy to construct functions $\Lambda$ which depend on
$\kkph$ and $|\kk_\perp|$ in a non-trivial manner. Hence, the Hamiltonian $\hat H$ is
invariant under non-trivial gauge transformations. On the other hand, it is also easy to construct
functions $\Lambda$ such that $\hat U_\Lambda$ does not commute with $\hat H$.

\section{Bound states}
\label{app:boundstate}

\subsection{The unperturbed vacuum}

The ground state of the non-interacting system is given  by
$\zeta_{\kkph,\kk}=|0,0,\emptyset\rangle$.
The two zeroes indicate the absence of horizontally, respectively
vertically polarized photons. The empty set indicates the absence of 
electrons and positrons.
This state is not an eigenstate of the interacting Hamiltonian $\hat H$.
One has
\be
[\hat H\zeta]_{\kkph,\kk}
&=&
[\hatHI\zeta]_{\kkph,\kk}\cr
&=&
\int_{\Ro^3}\upd \xx\,
[\hat A_\mu(x)\hat J^\mu(x)\zeta]_{\kkph,\kk}\bigg|_{x^0=0}\cr
&=&
\int_{\Ro^3}\upd \xx\,
A_{\mu,\kkph}(x)|0,0\rangle\,
\int\upd\kk'\,J^\mu_{\kk,\kk'}(x)|\emptyset\rangle\bigg|_{x^0=0}\cr
&=&
\frac {\lambda}{2\Nph (\kkph)}\frac {\uc c}{2(2\pi)^3}
\int_{\Ro^3}\upd \xx\,e^{i\kph_\mu x^\mu}\int\upd\kk'\,
e^{i(k_\nu+k'_\nu)x^\nu}\bigg|_{x^0=0}\cr
& &\times
\sum_\alpha\left[\varepsilon^{(H)}_\alpha(\kkph)|1,0\rangle
+\varepsilon^{(V)}_\alpha(\kkph)|0,1\rangle\right]\cr
& &
\times
\sum_{s=1,2}\sum_{t=3,4}\left[
\langle u^{(s)}(\kk)|\gamma^0\gamma^\alpha v^{(t)}(\kkp)\rangle
+\langle u^{(s)}(\kkp)|\gamma^0\gamma^\alpha v^{(t)}(\kk)\rangle\right]\cr
& &\times
\sigma_{s}^{(-)}\sigma_{t}^{(-)}|\emptyset\rangle\cr
&=&
-\frac {\lambda\uc c}{2\Nph (\kkph)}
\sum_\alpha\left[\varepsilon^{(H)}_\alpha(\kkph)|1,0\rangle
+\varepsilon^{(V)}_\alpha(\kkph)|0,1\rangle\right]\cr
& &
\times
\sum_{s=1,2}\sum_{t=3,4}
\langle u^{(s)}(\kk)|\gamma^0\gamma^\alpha v^{(t)}(-\kk-\kkph)\rangle
\,|\{s,t\}\rangle.
\nnee
The action of the Hamiltonian $\hat H$ on the free vacuum creates an electron/positron
pair together with a single photon.

The quantum expectation $\langle\zeta|\hat H\zeta\rangle$ of the energy vanishes.
An important question is whether there exist fields $\zeta$ in $\Gammaphel$ for which
$\langle\zeta|\hat H\zeta\rangle$ is negative, or even diverges to minus infinity.
This question is related to the problem of stability of matter and has
been studied extensively (See for instance the work of Lieb and coworkers \cite {LS09}).
A systematic study in the present context is postponed.
A partial answer is given in the following sections.

\subsection{Trial wave function}

Let us try to find a wave function $\zeta_{\kkph,\kk}$ for which the expectation value
$\langle\zeta|\hat H\zeta\rangle$ is strictly negative.

For all $\zeta$ of the form $\zeta_{\kkph,\kk}=\sum_{m,n}a_{m,n}(\kkph,\kk)|m,n,\emptyset\rangle$
is $\langle\zeta|\hat H\zeta\rangle\ge 0$. As a next step consider
wave functions of the form
\be
& &\zeta_{\kkph,\kk}=
\sqrt{\rhovac(\kkph,\kk)}|0,0,\emptyset\rangle\cr
& &\quad
+a_{0,0}(\kkph,\kk)|0,0,\{1\}\rangle
+a_{1,0}(\kkph,\kk)|1,0,\{1\}\rangle
+a_{0,1}(\kkph,\kk)|0,1,\{1\}\rangle.
\nnee
They describe a single electron with spin up, eventually accompanied by an electromagnetic wave
which is a superposition of a horizontally and a vertically polarized photon.
The superposition with a wave function with vanishing electron/positron field is needed
in order to satisfy the conflicting requirements of proper normalization and of a finite
quantum expectation of the energy.
Proper normalization requires that for all $\kkph,\kk$
\be
1&=&\rhovac(\kkph,\kk)+|a_{0,0}(\kkph,\kk)|^2+|a_{1,0}(\kkph,\kk)|^2+|a_{0,1}(\kkph,\kk)|^2.
\nnee
Finiteness of the total energy requires that the density of the vacuum $\rhovac(\kkph,\kk)$
tends to 1 fast enough when $|\kkph|$ and $|\kk|$ become large.

The kinetic energy of the fields equals
\be
\calEph&=&\lu^3\int\upd\kkph\,\hbar c|\kkph|\rhoph(\kkph),\cr
\calEel&=&\lu^3\int\upd\kk\,\hbar \omega(\kk)\rhoel(\kk),
\nnee
with
\be
\rhoph(\kkph)&=&\lu^3\int\upd\kk\,
\left[
|a_{1,0}(\kkph,\kk)|^2+|a_{0,1}(\kkph,\kk)|^2)\right],\cr
\rhoel(\kk)&=&\lu^3\int\upd\kkph\,
\left[
|a_{0,0}(\kkph,\kk)|^2+|a_{1,0}(\kkph,\kk)|^2+|a_{0,1}(\kkph,\kk)|^2)\right]\cr
&=&
\lu^3\int\upd\kkph\,\left[1-\rhovac(\kkph,\kk)\right].
\nnee

Before evaluating the interaction energy first consider
\be
& &[\hat \HI\zeta]_{\kkph,\kk}\cr
&=&
\int\upd\xx\,\sum_\alpha A_{\alpha,\kkph}(x)|0,0\rangle\cr
& &
\times \int\upd\kkp\,J_{\alpha,\kk,\kk'}(x) a_{0,0}(\kkph,\kkp)|\{1\}\rangle\bigg|_{x^0=0}\cr
& &
+\int\upd\xx\,\sum_\alpha A_{\alpha,\kkph}(x)|1,0\rangle\cr
& &
\times \int\upd\kkp\,J_{\alpha,\kk,\kk'}(x) a_{1,0}(\kkph,\kkp)|\{1\}\rangle\bigg|_{x^0=0}\cr
& &
+\int\upd\xx\,\sum_\alpha A_{\alpha,\kkph}(x)|0,1\rangle\cr
& &
\times \int\upd\kkp\,J_{\alpha,\kk,\kk'}(x) a_{0,1}(\kkph,\kkp)|\{1\}\rangle\bigg|_{x^0=0}\cr
&=&
\frac {\lambda\uc c}{4\Nph (\kkph)(2\pi)^3}\int\upd\xx\,\sum_\alpha 
\left[\varepsilon^{(H)}_\alpha(\kkph)e^{-i\kkph\cdot\xx}\ahdagger
+\varepsilon^{(V)}_\alpha(\kkph)e^{-i\kkph\cdot\xx}\avdagger\right]
|0,0\rangle\cr
& &
\times \int\upd\kkp\,
\left[e^{-i(\kk-\kkp)\cdot\xx}
\langle u^{(1)}(\kk)|\gamma^0\gamma^\alpha u^{(1)}(\kkp)\rangle
+e^{i(\kk-\kkp)\cdot\xx}
\langle u^{(1)}(\kkp)|\gamma^0\gamma^\alpha u^{(1)}(\kk)\rangle\right]\cr
& &\qquad\times
a_{0,0}(\kkph,\kkp)|\{1\}\rangle\cr
& &
+\frac {\lambda\uc c}{4\Nph (\kkph)(2\pi)^3}\int\upd\xx\,\sum_\alpha 
\varepsilon^{(H)}_\alpha(\kkph)e^{i\kkph\cdot\xx}\ah|1,0\rangle\cr
& &
\times \int\upd\kkp\,
\left[e^{-i(\kk-\kkp)\cdot\xx}
\langle u^{(1)}(\kk)|\gamma^0\gamma^\alpha u^{(1)}(\kkp)\rangle
+e^{i(\kk-\kkp)\cdot\xx}
\langle u^{(1)}(\kkp)|\gamma^0\gamma^\alpha u^{(1)}(\kk)\rangle\right]\cr
& &\qquad\times
a_{1,0}(\kkph,\kkp)|\{1\}\rangle\cr
& &
+\frac {\lambda\uc c}{4\Nph (\kkph)(2\pi)^3}\int\upd\xx\,\sum_\alpha 
\varepsilon^{(V)}_\alpha(\kkph)e^{i\kkph\cdot\xx}\av|0,1\rangle\cr
& &
\times \int\upd\kkp\,
\left[e^{-i(\kk-\kkp)\cdot\xx}
\langle u^{(1)}(\kk)|\gamma^0\gamma^\alpha u^{(1)}(\kkp)\rangle
+e^{i(\kk-\kkp)\cdot\xx}
\langle u^{(1)}(\kkp)|\gamma^0\gamma^\alpha u^{(1)}(\kk)\rangle\right]\cr
& &\qquad\times
a_{0,1}(\kkph,\kkp)|\{1\}\rangle\cr
& &
+\cdots.
\nnee
The omitted terms are orthogonal to $|0,0,\emptyset\rangle$ and $|m,n,\{1\}\rangle$.
Integration over $\xx$ gives
\be
& &
[\hat \HI\zeta]_{\kkph,\kk}\cr
&=&
\frac {\lambda\uc c}{4\Nph (\kkph)}\sum_\alpha 
\left[
\varepsilon^{(H)}_\alpha(\kkph)|1,0\rangle
+\varepsilon^{(V)}_\alpha(\kkph)|0,1\rangle
\right]\cr
& &
\times \int\upd\kkp\,
\bigg[\delta(-\kkph-\kk+\kkp)
\langle u^{(1)}(\kk)|\gamma^0\gamma^\alpha u^{(1)}(\kkp)\rangle\cr
& &\quad
+\delta(-\kkph+\kk-\kkp)
\langle u^{(1)}(\kkp)|\gamma^0\gamma^\alpha u^{(1)}(\kk)\rangle\bigg]
a_{0,0}(\kkph,\kkp)|\{1\}\rangle\cr
& &
+\frac {\lambda\uc c}{4\Nph (\kkph)}\sum_\alpha 
\varepsilon^{(H)}_\alpha(\kkph)|0,0\rangle\cr
& &
\times \int\upd\kkp\,
\bigg[\delta(\kkph-\kk+\kkp)
\langle u^{(1)}(\kk)|\gamma^0\gamma^\alpha u^{(1)}(\kkp)\rangle\cr
& &\quad
+\delta(\kkph+\kk-\kkp)
\langle u^{(1)}(\kkp)|\gamma^0\gamma^\alpha u^{(1)}(\kk)\rangle\bigg]
a_{1,0}(\kkph,\kkp)|\{1\}\rangle\cr
& &
+\frac {\lambda\uc c}{4\Nph (\kkph)}\sum_\alpha 
\varepsilon^{(V)}_\alpha(\kkph)|0,0\rangle\cr
& &
\times \int\upd\kkp\,
\bigg[\delta(\kkph-\kk+\kkp)
\langle u^{(1)}(\kk)|\gamma^0\gamma^\alpha u^{(1)}(\kkp)\rangle\cr
& &\quad
+\delta(\kkph+\kk-\kkp)
\langle u^{(1)}(\kkp)|\gamma^0\gamma^\alpha u^{(1)}(\kk)\rangle\bigg]
a_{0,1}(\kkph,\kkp)|\{1\}\rangle\cr
& &
+\cdots\cr
&=&
\frac {\lambda\uc c}{4\Nph (\kkph)}\sum_\alpha 
\left[
\varepsilon^{(H)}_\alpha(\kkph)|1,0,\{1\}\rangle
+\varepsilon^{(V)}_\alpha(\kkph)|0,1,\{1\}\rangle\right]\cr
& &
\times 
\bigg[
\langle u^{(1)}(\kk)|\gamma^0\gamma^\alpha u^{(1)}(\kk+\kkph)\rangle a_{0,0}(\kkph,\kk+\kkph)\cr
& &\quad
+
\langle u^{(1)}(\kk-\kkph)|\gamma^0\gamma^\alpha u^{(1)}(\kk)\rangle a_{0,0}(\kkph,\kk-\kkph)\bigg]\cr
& &
+\frac {\lambda\uc c}{4\Nph (\kkph)}\sum_\alpha 
\varepsilon^{(H)}_\alpha(\kkph)|0,0,\{1\}\rangle\cr
& &
\times 
\bigg[
\langle u^{(1)}(\kk)|\gamma^0\gamma^\alpha u^{(1)}(\kk-\kkph)\rangle a_{1,0}(\kkph,\kk-\kkph)\cr
& &\quad
+
\langle u^{(1)}(\kk+\kkph)|\gamma^0\gamma^\alpha u^{(1)}(\kk)\rangle a_{1,0}(\kkph,\kk+\kkph)\bigg]\cr
& &
+\frac {\lambda\uc c}{4\Nph (\kkph)}\sum_\alpha 
\varepsilon^{(V)}_\alpha(\kkph)|0,0,\{1\}\rangle\cr
& &
\times 
\bigg[
\langle u^{(1)}(\kk)|\gamma^0\gamma^\alpha u^{(1)}(\kk-\kkph)\rangle a_{0,1}(\kkph,\kk-\kkph)\cr
& &\quad
+
\langle u^{(1)}(\kk+\kkph)|\gamma^0\gamma^\alpha u^{(1)}(\kk)\rangle a_{0,1}(\kkph,\kk+\kkph)\bigg]\cr
& &
+\cdots.
\nnee
The quantum expectation of the interaction energy becomes
\be
\calEint
&=&-\lu^3\int\upd\kkph\,\frac {\lambda\uc c}{4\Nph (\kkph)}\sum_\alpha \cr
& &
\times\left[
\varepsilon^{(H)}_\alpha(\kkph) w^{(H)}_\alpha(\kkph)
+
\varepsilon^{(V)}_\alpha(\kkph) w^{(V)}_\alpha(\kkph)\right]
\nnee
with
\be
w^{(H)}_\alpha(\kkph)
&=&
-2\Re \lu^3\int\upd\kk\,\overline{a_{1,0}(\kkph,\kk)}a_{0,0}(\kkph,\kk+\kkph)\cr
& &
\quad\times 
\langle u^{(1)}(\kk)|\gamma^0\gamma^\alpha u^{(1)}(\kk+\kkph)\rangle 
\nnee
and
\be
w^{(V)}_\alpha(\kkph)
&=&
-2\Re \lu^3\int\upd\kk\,\overline{a_{0,1}(\kkph,\kk)}a_{0,0}(\kkph,\kk+\kkph)\cr
& &
\quad\times 
\langle u^{(1)}(\kk)|\gamma^0\gamma^\alpha u^{(1)}(\kk+\kkph)\rangle.
\nnee

The terms which contribute describe the 
interaction of the photon with the spin of the electron.

\subsection{Variational approach}
\label{app:var}

Consider now the problem of minimizing the total energy given a fixed value
for the density of the vacuum $\rhovac(\kkph,\kk)$.
Variation of $\overline{a_{1,0}(\kkph,\kk)}$ gives
\be
a_{1,0}(\kkph,\kk)&=&-U^{(H)}(\kkph,\kk)a_{0,0}(\kkph,\kk+\kkph)
\nnee
with
\be
U^{(H/V)}(\kkph,\kk)
&=&
\frac{\lambda\uc }
{4\Nph (\kkph)\hbar |\kkph|}
\sum_\alpha 
\varepsilon^{(H/V)}_\alpha(\kkph) 
\langle u^{(1)}(\kk)|\gamma^0\gamma^\alpha u^{(1)}(\kk+\kkph)\rangle.
\nnee
Similarly,
\be
a_{0,1}(\kkph,\kk)&=&-U^{(V)}(\kkph,\kk)a_{0,0}(\kkph,\kk+\kkph).
\nnee
The normalization condition becomes
\be
& &
1=\rhovac(\kkph,\kk)+|a_{0,0}(\kkph,\kk)|^2
+U_\perp^2(\kkph,\kk)\,|a_{0,0}(\kkph,\kk+\kkph)|^2,\quad
\label{int:var:normcond}
\ee
with
\be
U_\perp^2(\kkph,\kk)&=&|U^{(H)}(\kkph,\kk)|^2+|U^{(V)}(\kkph,\kk)|^2.
\nnee
The functions $w^{(H)}_\alpha(\kkph)$ and $w^{(V)}_\alpha(\kkph)$ are of the form
\be
w^{(H/V)}_\alpha(\kkph)&=&2\Re\lu^3\int\upd\kk\,\overline{U^{(H/V)}(\kkph,\kk)}
|a_{0,0}(\kkph,\kk+\kkph)|^2\cr
& &
\times
\langle u^{(1)}(\kk)|\gamma^0\gamma^\alpha u^{(1)}(\kk+\kkph)\rangle.
\nnee
The interaction energy becomes
\be
\calEint
&=&-2\lu^6\int\upd\kkph\,\int\upd\kk\,\frac {\lambda\uc c}{4\Nph (\kkph)}|a_{0,0}(\kkph,\kk+\kkph)|^2 \cr
& &
\times
\sum_\alpha\Re\left[
\varepsilon^{(H)}_\alpha(\kkph) \overline{U^{(H)}(\kkph,\kk)}
+
\varepsilon^{(V)}_\alpha(\kkph) \overline{U^{(V)}(\kkph,\kk)}\right]\cr
& &
\quad\times 
\langle u^{(1)}(\kk)|\gamma^0\gamma^\alpha u^{(1)}(\kk+\kkph)\rangle\cr
&=&
-2\lu^6\int\upd\kkph\,\hbar c|\kkph|\int\upd\kk\,|a_{0,0}(\kkph,\kk+\kkph)|^2 U^2_\perp(\kkph,\kk)\cr
&=&
-2\calEph.
\nnee
The interaction energy is minus twice the kinetic energy of the photon field.
This result is typical for a quadratic minimization problem. During the
minimization the energy of the electron field is kept constant.
Hence one can conclude that for an electron field with a given kinetic energy
and no photons present there always exists an interacting system
where the energy spectrum of the electron field is unmodified but the total
energy is lowered by adding the photon field.

For further use, note that
the kinetic energy of the photon field can be written as
\be
\calEph
&=&
\lu^6\int\upd\kkph\,\hbar c|\kkph|\int\upd\kk\,
|a_{0,0}(\kkph,\kk)|^2
U^2_\perp(\kkph,\kk-\kkph).\cr
& &
\label{bound:epf}
\ee
The density of the electron field equals 
\be
\rhoel(\kk)&=&\lu^3\int\upd\kkph\,
\left[
|a_{0,0}(\kkph,\kk)|^2+|a_{1,0}(\kkph,\kk)|^2+|a_{0,1}(\kkph,\kk)|^2)\right]\cr
&=&
\lu^3\int\upd\kkph\,
\left[|a_{0,0}(\kkph,\kk)|^2
1+U^2_\perp(\kkph,\kk)|a_{0,0}(\kkph,\kk+\kkph)|^2\right].\cr
& &
\label{bound:def}
\ee

\subsection{Long-wavelength analysis}
\label{app:longwave}
Above it is shown that for a wave function of the form
\be
\zeta_{\kkph,\kk}&=&
\sqrt{\rhovac(\kkph,\kk)}|0,0,\emptyset\rangle
+a_{0,0}(\kkph,\kk)|0,0,\{1\}\rangle\cr
& &\quad
-U^{(H)}(\kkph,\kk)a_{0,0}(\kkph,\kk+\kkph)|1,0,\{1\}\rangle\cr
& &\quad
-U^{(V)}(\kkph,\kk)a_{0,0}(\kkph,\kk+\kkph)|0,1,\{1\}\rangle
\nnee
the interaction energy is minus twice the kinetic energy of the photon field.
By explicit calculation one shows that coefficients $a_{0,0}(\kkph,\kk)$
exist such that the wave function is physically acceptable.
These calculations lead to
\be
U^{(H/V)}(\kkph,\kk)
&=&
-\frac{\lambda\uc c}
{4\Nph (\kkph)\hbar |\kkph|}\,\sum_\alpha \varepsilon^{(H/V)}_\alpha(\kkph)k_\alpha\cr
& &\times
\frac{\omega(\kk)+\omega(\kk+\kkph)+2c\mstar}
{\sqrt{\omega(\kk+\kkph)[\omega(\kk+\kkph)+c\mstar]}\,{\sqrt{\omega(\kk)[\omega(\kk)+c\mstar]}}}.\cr
& &
\label{lwa:UHVexpr}
\ee
This gives
\be
U^2_\perp(\kkph,\kk)
&=&
\left(\frac{\lambda\uc c}
{4\Nph (\kkph)\hbar |\kkph|}\right)^2\,
\left(|\kk|^2-\frac{(\kk\cdot\kkph)^2}{|\kkph|^2}\right)\cr
& &\times
\frac{[\omega(\kk)+\omega(\kk+\kkph)+2c\mstar]^2}
{\omega(\kk+\kkph)[\omega(\kk+\kkph)+c\mstar]\,\omega(\kk)[\omega(\kk)+c\mstar]}\cr
&\ge&
\left(\frac{\lambda\uc c}
{4\Nph (\kkph)\hbar |\kkph|}\right)^2\,
\left(|\kk|^2-\frac{(\kk\cdot\kkph)^2}{|\kkph|^2}\right)
\frac{4}{\omega(\kk)\omega(\kk+\kkph)}.
\nnee

A long wavelength approximation is
\be
U^2_\perp(\kkph,\kk)
&=&
\left(\frac{\lambda\uc c}{4\Nph (\kkph)\hbar |\kkph|}\right)^2\,
\left(|\kk|^2-\frac{(\kk\cdot\kkph)^2}{|\kkph|^2}\right)\cr
& &\times
\frac{4}{\omega^2(\kk)}
\left[1-\frac{\kk\cdot\kkph}{\mstar^2+|\kk|^2}+\mbox{O}(|\kkph|^2)\right].
\nnee
If $\kkph$ and $\kk$ are not parallel then this expression diverges as $|\kkph|^{-3}$.
From the normalization condition (\ref {int:var:normcond}) then follows that
$a_{0,0}(\kkph,\kk)$ should vanish in the long-wavelength limit as $|\kkph|^{3}$
or it should vanish in all regions away from the longitudinal direction.
This observation leaves room for two types of solutions.

\section{Emergent Coulomb forces}
\label{app:emerg}

In standard QED the electromagnetic field is described by 4 independent
operators $\hat A_\mu(x)$. In the present approach one component, namely $\hat A_0(x)$
is identically zero and the 3 remaining components $\hat A_\alpha(x)$, $\alpha=1,2,3$,
satisfy the orthogonality relation
\be
\sum_\alpha \kph_\alpha A_{\alpha,\kkph}(x)&=&0.
\nnee
Hence, the electromagnetic quantum field, as treated in the present work,
has only two degrees of freedom. In particular, the electric field operators 
$\hat E_\alpha(x)$ satisfy Gauss' law in absence of charges (See (\ref {photon:gauss})).
This can be justified with the argument that the full law, including a
source term in the r.h.s., will
emerge after the interaction with the electron/positron field is taken into account.
This argument is supported by the existence \cite{NJ17}
of a transformation
of the field operators $\hat E_\alpha(x)$ which maps the homogeneous law
of Gauss onto the full version of the law.

Introduce new electric field operators
\be
\hat E''_\alpha(x)&=&\hat E'_\alpha(x)\cr
&+&\frac{\mu_0 c}{4\pi}\frac{\partial\,}{\partial x^\alpha}\int\upd\yy\,
\frac{1}{|\xx-\yy|}\cr
& &\times \hat U(-x^0)\hat j^0(\yy,0)\hat U(x^0).\cr
& &
\label{emerg:def2}
\ee
Here, $\hat U(x^0)=\exp(-ix^0\hat H/\hbar c)$ is the time evolution of the interacting system.
The new operators are marked with a double prime to distinguish them from the
operators of the non-interacting system and those of the interacting system. The latter are denoted
with a single prime. One verifies immediately that
Gauss' law is satisfied
\be
\sum_\alpha\frac{\partial\,}{\partial x^\alpha}\hat E''_\alpha(x)
&=&
-\mu_0 c\,\hat j^{0\prime}(x).
\nnee

The second term of (\ref {emerg:def2}) is the Coulomb contribution to the electric field.
The curl of this term vanishes. Hence it is obvious to take
\be
\hat B''_\alpha(x)&\equiv& \hat B'_\alpha(x).
\nnee
This implies the second of the four equations of Maxwell,
stating that the divergence of $\hat B''_\alpha(x)$ vanishes.
In addition, the fourth equation, absence of magnetic charges,
follows immediately because $\hat E''(x)$ and $\hat E'(x)$
have the same curl. Remains Faraday's law to be written as
\be
(\nabla\times\hat B''(x))_\alpha-\frac 1{c}\frac{\partial\,}{\partial x^0}\hat E''_\alpha(x)
&=&
-\mu_0\,\hat j''_\alpha(x)
\nnee
with
\be
\hat j''_\alpha(x)
&=&
-\frac 1{\mu_0 c}\frac{\partial\,}{\partial x^0}\left(\hat E''_\alpha(x)-\hat E'_\alpha(x)\right).
\nnee
Finally, take $\hat j''_0(x)=\hat j'_0(x)$.
A short calculation shows that the newly defined current operators $\hat j''_\mu(x)$
satisfy the continuity equation.

One concludes that a formalism of QED is possible which does not postulate the
existence of longitudinal or scalar photons. Two pictures coexist: the original
Heisenberg picture and what is called here the {\em emergent} picture.
In both pictures the time evolution of all operators is the same, but
the definition of the electromagnetic field operators differs.
In the original description only transversely polarized photons exist.
On the other hand, the field operators of the emergent
picture satisfy the full Maxwell equations, including Coulomb forces.



\end{document}